\newtheorem{proposition}{Proposition}

\documentclass[lettersize,journal]{IEEEtran}

\usepackage{amsmath,amssymb,amsfonts}

\usepackage{newtxtext,newtxmath}

\usepackage[utf8]{inputenc} 
\usepackage[T1]{fontenc}    
\usepackage{textcomp}       

\usepackage{graphicx}
\usepackage[caption=false,font=normalsize,labelfont=sf,textfont=sf]{subfig}
\usepackage{booktabs}
\usepackage{threeparttable}
\usepackage{array}
\usepackage{colortbl}
\usepackage{multicol}
\usepackage{stfloats}

\usepackage[dvipsnames]{xcolor}

\usepackage{algorithm}
\usepackage{algorithmic}

\usepackage{verbatim}
\usepackage{url}
\usepackage{balance}

\usepackage{fancyhdr}
\def\BibTeX{{\rm B\kern-.05em{\sc i\kern-.025em b}\kern-.08em
		    T\kern-.1667em\lower.7ex\hbox{E}\kern-.125emX}}
\usepackage{balance}
\usepackage{cite}
\begin{document}
\title{Continuous-Variable  MIMO THz Quantum Secret Sharing:  Gaussian-modulation  and Passive-modulation}
\author{Leixin Wu, Jiayu Pan, Fangzhe Chen, Lingtao Zhang, Bowen Zheng, and Tie Qiu
\thanks{This work is supported by National Natural Science Foundation of China 62402435, by Ningbo Yongjiang Talent Programme 2023A-398-G, and Natural Science Foundation of Ningbo 2024J205.
	\textit{(Corresponding author: Jiayu Pan.)}
	
	Leixin Wu and Lingtao Zhang are with the College of Electronic Information and Physics, Central South University of Forestry and Technology, Changsha {\rm 410004}, China (e-mail: 20241100491@csuft.edu.cn; zhang@csuft.edu.cn). 
		
		Jiayu Pan and Tie Qiu are with the School of Computer Science and Engineering, Northeastern University,  Shenyang {\rm 110819}, China (e-mail: panjiayu@cse.neu.edu.cn; qiutie@ieee.org).
		
		Fangzhe Chen is with the School of Software Technology , Zhejiang University,
		Hangzhou {\rm 310058}, China (e-mail: 22451317@zju.edu.cn).
		
		Bowen Zheng is with the School and Hospital of Stomatology, China Medical University, Shenyang 110002, China (e-mail: bwzheng@cmu.edu.cn).
} }

\markboth{}%
{How to Use the IEEEtran \LaTeX \ Templates}

\maketitle

\begin{abstract}
	Although QKD enables information-theoretically secure key distribution, it is primarily designed for point-to-point communication and cannot directly support multi-user collaborative scenarios. To address this limitation, quantum secret sharing (QSS) has been proposed to enable secure multiparty communication.  However, current QSS protocols employ a single-input single-output (SISO) channel  for transmission, which severely constrains the achievable secret key rate (SKR) and transmission distance. To meet the demand for high SKR transmission in 6G wireless communication, this paper proposes a continuous-variable (CV) QSS protocol based on a multiple-input multiple-output (MIMO) architecture  operating in the terahertz (THz) band.  In this scheme, transmit–receive beamforming is employed to decompose the MIMO channel into multiple single-input single-output (SISO) subchannels, enabling parallel signal transmission and increasing both the SKR and transmission distance.  We describe the QSS transmission protocol and derive the SKR expressions for eight QSS protocol variants {under Gaussian collective attacks.} At the transmitter side, we consider two modulation schemes: Gaussian modulation and passive modulation, which are the two most commonly used schemes to generate coherent states. At the receiver side, we consider two detection schemes: homodyne and heterodyne detection. In addition, we derive two versions of the SKR: asymptotic and { composable} finite-size, which quantify the upper bounds of the SKR and {the achievable performance under finite resources}, respectively. {Simulation results demonstrate that, under ideal assumptions (including perfect channel state information, perfect phase synchronization, and ideal beamforming), the Gaussian modulation protocol with a $32\times32$ antenna configuration and the passive modulation protocol with a $1024\times1024$ antenna configuration achieve transmission distances of 14.99 m and 160 m in the atmospheric channel, respectively. } {These results provide an idealized theoretical benchmark for evaluating the potential performance gains of MIMO-assisted THz CV-QSS in indoor and short-range outdoor wireless networks.}

\end{abstract}

\begin{IEEEkeywords}
 Continuous-variable quantum secret sharing,  Multiple-input multiple-output, { Composable} finite-size analysis, Terahertz, Gaussian-modulation coherent state, Passive-modulation coherent state, Wireless communication network
\end{IEEEkeywords}

\section{INTRODUCTION}
\IEEEPARstart{T}{he} sixth-generation (6G) mobile networks bring unprecedented demands for ultra-high data rates, ultra-low latency, massive connectivity, and robust security \cite{1,2,3,4}. These requirements are vital for enabling advanced applications such as holographic immersive communication \cite{5}, ultra-reliable low-latency communication (URLLC) \cite{6}, massive machine-type communication (mMTC) \cite{7}, and integrated sensing and communication (ISAC) \cite{8}. To meet these demands, there are increasing investigations on higher-frequency communication technologies, with the terahertz (THz) band recognized as a key enabler for future wireless systems \cite{9,10}.

{THz communication provides abundant and largely untapped spectrum resources and supports significantly higher data rates than millimeter-wave systems \cite{11}. Owing to its short wavelength, THz communication can exploit highly directional beamforming and spatial multiplexing techniques, making it well suited for high-capacity short-range wireless communications \cite{12}. Although the highly directional nature of THz beams can reduce signal leakage compared with conventional radio-frequency systems, it does not guarantee information-theoretic security. In practice, confidential information may still be exposed through beam scattering, diffraction, reflections, or passive beam-splitting attacks, allowing an eavesdropper to obtain part of the transmitted signal without introducing detectable disturbances \cite{13,14,15}. Therefore, ensuring information-theoretic security remains a fundamental challenge for THz communication systems.}

Quantum key distribution (QKD) leverages the principles of quantum mechanics to establish unconditionally secure keys over insecure channels \cite{16,17}. {QKD derives its information-theoretic security from fundamental principles of quantum mechanics, including the no-cloning theorem and quantum uncertainty, together with rigorous security proofs \cite{18,19,20}.} QKD can be divided into two main types: discrete-variable (DV) and continuous-variable (CV) \cite{21}. {DV-QKD typically relies on single-photon or weak coherent states, has achieved long-distance transmission in many implementations, and has already been commercialized \cite{22,23,24}.} In contrast, {CV-QKD remains largely in the experimental and theoretical stage but is highly compatible with existing fiber-optic and telecommunication infrastructure} \cite{25,26}. Moreover, CV-QKD supports channel multiplexing, enabling higher SKRs. {These advantages make CV-QKD a promising candidate for future high-speed, short-range, broadband quantum communications} \cite{27}.

To meet the ultra-high bandwidth and low-latency requirements of 6G networks, CV-QKD has been investigated in the THz band \cite{28,29}. {THz CV-QKD uses narrow, highly directional beams, which can relax the stringent optical-alignment requirements in free-space quantum communication \cite{30}.} More importantly, unlike conventional THz communication systems, THz CV-QKD can detect potential eavesdropping attempts through the disturbance introduced by quantum measurements, thereby enabling information-theoretically secure key distribution \cite{31}. These features make THz CV-QKD suitable for indoor and short-range quantum communications.

However, conventional QKD systems are limited to point-to-point key exchange and thus cannot support the secure multicasting and multi-user collaboration required in 6G networks \cite{32}. To overcome this limitation, a continuous-variable quantum secret sharing (CV-QSS) protocol based on a $(k, n)$ threshold structure has been proposed \cite{33,34,35,36}. In this protocol, the key dealer shares a secret key with $n$ users, and decryption of the ciphertext requires the collaboration of $k$ users, where $k \leq n$. {This setup ensures that any subset of fewer than $k$ users cannot decrypt the message independently.} The QSS scheme prevents unauthorized access to the secret key, thereby maintaining the confidentiality of the communication. This paper focuses on the $(n, n)$-threshold case, where all users must participate.

Within CV-QSS, two modulation schemes are commonly considered: Gaussian modulation coherent state (GMCS) and passive modulation coherent state (PMCS) \cite{37,38}. Gaussian modulation encodes information onto coherent states with Gaussian-distributed amplitude and phase, achieving asymptotic optimal performance but requiring high-speed, high-precision modulators \cite{37}.  Passive modulation, by contrast, generates states from thermal noise using  attenuators and heterodyne detectors, eliminating the need for active high-speed modulation and simplifying system implementation \cite{38}. 

However, existing QSS protocols rely on single-input single-output (SISO) channels for transmission. This substantially limits both the achievable SKR and the maximum communication distance.  Inspired by the success of multiple-input multiple-output (MIMO) systems in classical communication and QKD systems \cite{9,388,389}, we propose a MIMO THz CV-QSS protocol for secure free-space communication to overcome the limitations of conventional SISO schemes. In this scheme, the MIMO architecture provides multiple spatial eigenchannels, enabling parallel secret-key share generation within each protocol round. This technology not only increases the transmission distance but also enhances the overall SKR.  {Here, the SKR is measured in bit/use and refers to the minimum net number of information-theoretically secure key bits shared between the dealer and any authorized user during one MIMO-QSS protocol round, where one use includes one simultaneous use of all parallel SISO eigenchannels. } Moreover, by leveraging the ultra-high bandwidth and strong beam directivity of THz communication, the proposed QSS protocol achieves reliable multi-user key distribution over free-space channels while maintaining rigorous security, making it a promising candidate for secure and collaborative 6G networks. Furthermore, both GMCS and PMCS schemes are investigated to balance performance and implementation feasibility. The main contributions of this paper are summarized as follows:
\begin{itemize}
\item[1)]We propose a MIMO-assisted CV-QSS protocol in the THz band, {where secure communication between the dealer and the users is established through multi-user cooperation and multi-hop forwarding. Compared with conventional MIMO CV-QKD protocols \cite{9}, beamforming  must be performed sequentially over multiple hops. Consequently, the achievable SKR is jointly constrained by the number of available eigenchannels across all hops.}
\item[2)]We describe the complete communication procedure of the proposed MIMO THz CV-QSS scheme, which consists of two main phases: quantum transmission and classical post-processing. 
\item[3)]We derive the SKR expressions for eight QSS protocol variants {under Gaussian collective attacks}, which employ GMCS or PMCS at the transmitter and homodyne or heterodyne detection at the receiver. {Compared with previous THz CV-QSS protocols, which are commonly analyzed using security frameworks originally developed for idealized pure-state CV-QKD systems \cite{29}, we explicitly account for the mixed-state nature induced by thermal photons and incorporate thermal noise into both the performance and security analysis, thereby providing a more realistic and physically consistent evaluation framework.}  Furthermore, we perform both asymptotic and { composable} finite-size analyses to evaluate the theoretical upper bounds of the SKRs and the composable SKR in finite-resource scenarios.
\item[4)]{We conduct numerical simulations to evaluate the theoretical performance trends of the proposed protocol over the atmospheric channel. Under the adopted idealized assumptions, the proposed MIMO-assisted CV-QSS scheme can improve the SKRs and achievable transmission distances compared with conventional SISO-QSS schemes.} Note that we have compared under various conditions, {including both PMCS and GMCS state preparation schemes and both homodyne and heterodyne measurement strategies employed by the dealer.}
\end{itemize}
Notation: $\mathbf{1}_{M\times N}$ and $\mathbf{0}_{M\times N}\in\mathbb{C}^{M\times N}$ denote the $M\times N$ matrices whose elements are all $1$ and $0$, respectively. The notation $\mathbf{A}^{\dagger}$ refers to the conjugate transpose of $\mathbf{A}$, while $\mathcal{N}(M,N)$ denotes a Gaussian distribution with mean $M$ and variance $N$, and $\chi^2$ denotes a chi-square distribution. We denote by $\mathbf{I}_k$ the $k\times k$ identity matrix, and the Pauli matrix $\mathbf{Z}$ as $\mathbf{Z}=\operatorname{diag}\left\{1,-1\right\}$.
\begin{figure*}[!t]
	\centering
	\includegraphics[width=\linewidth]{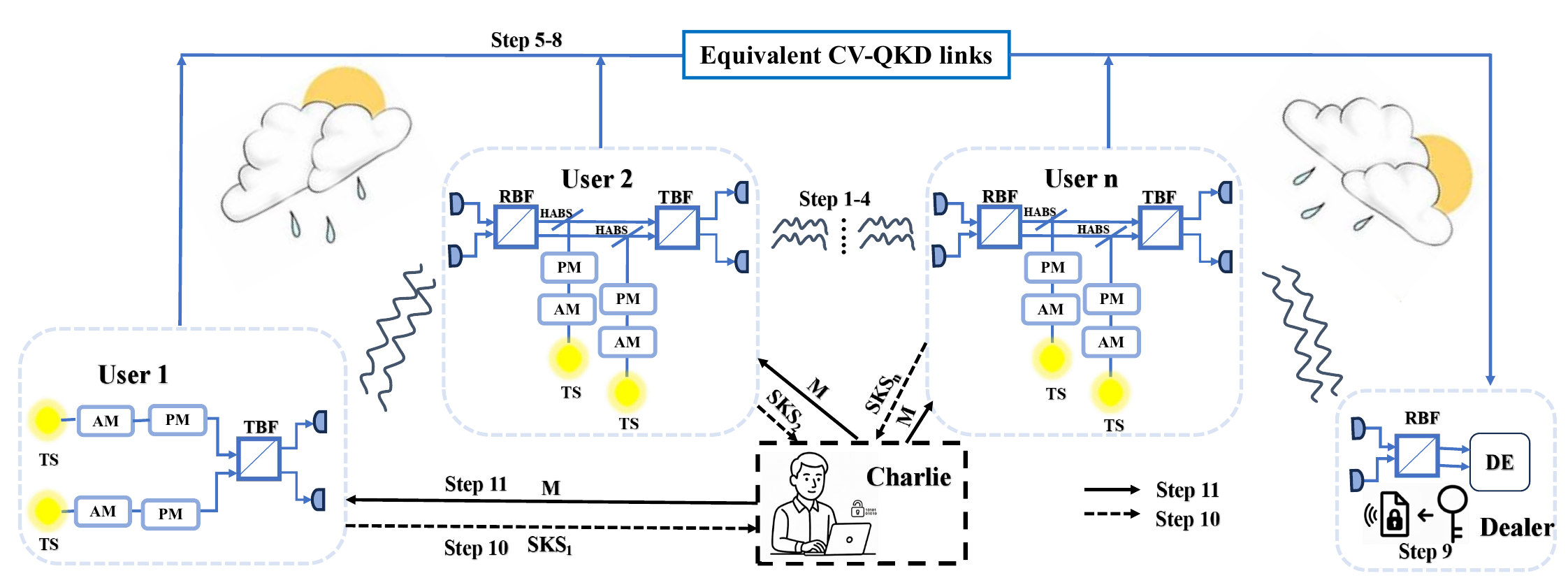}
	\caption{Schematic diagram of the GMCS MIMO THz-QSS protocol. AM and PM: amplitude modulator and phase modulator, TBF and RBF: transmit and receive beamforming, DE: detection, TS: terahertz source, SKS: secret-key share.}
	\label{fig1}
\end{figure*}
\section{SYSTEM MODEL}
This paper considers $(n, n)$-threshold GMCS and PMCS CV-QSS protocols, in which all users must collaborate to reconstruct the final secret and recover the plaintext message. It is worth noting that, for simplicity, the dealer is treated as the $(n+1)$-th user in this paper. {Throughout this work, the dealer is assumed to be trusted, similar to the trusted-receiver setting commonly adopted in conventional CV-QKD protocols.} The atmospheric channel matrix model is given in Appendix \ref{appendixa}. 
\subsection{GMCS scheme}
Figure~\ref{fig1} illustrates the system model of the MIMO THz QSS protocol based on GMCS, where transmit and receive beamforming are employed to decompose the MIMO channel into multiple parallel SISO subchannels. In this model, the $k$-th user communicates with the subsequent $(k+1)$-th user via a THz wireless channel. To enable uninterrupted signal transmission, all intermediate users $h$ (except user 1 and the dealer) are equipped with $N_{t_h}$ transmit antennas and $N_{r_h}$ receive antennas. This enables all the intermediate users to receive modes from the preceding user, apply the  beamforming and coupling operations, and forward the resulting modes to the next user. User 1 only transmits signals and is thus equipped  with $N_{t_1}$ transmit antennas, while the dealer only receives signals and uses $N_{r_D}$ receive antennas. {During the classical post-processing phase, we introduce an auxiliary trusted entity, Charlie, who  participates only in the classical post-processing stage. He temporarily collects the secret shares transmitted by legitimate users through secure channels and reconstructs the plaintext message, without participating in quantum-state preparation, transmission, or measurement. Therefore, unlike conventional trusted relays, Charlie does not access or manipulate quantum states and introduces no additional quantum attack surface. Moreover, Charlie does not store any long-term secret information, and its functionality can be replaced by a fully distributed reconstruction process or further distributed among multiple semi-trusted nodes, thereby avoiding a single point of failure.} { In practice, the authenticity of Charlie and the integrity of the classical communication with Charlie should be guaranteed by authenticated classical channels or information-theoretic message authentication mechanisms. For security-critical applications, Charlie can be removed and replaced by a distributed reconstruction process among the authorized users. In this case, the secret-key shares are transmitted to a selected target user through secure authenticated channels, and the plaintext is reconstructed only by this target user. Therefore, this distributed reconstruction mode is more suitable for secure one-to-one transmission between the dealer and the selected target user. By contrast, the trusted-combiner-based reconstruction adopted in this work provides a convenient implementation for multi-user plaintext recovery.} 
 The communication procedure consists of two main phases: quantum transmission and  classical post-processing. The detailed steps of the protocol are as follows:

\noindent $Quantum$ $transmission$
\begin{itemize}
\item \emph{Step 1:}	User 1 generates $r_{\mathrm{min}} = \min\{r_1, r_2, \ldots, r_n\}$ thermal states using a THz source. After applying amplitude modulator (AM) and phase modulator (PM), these states are prepared as GMCS, described as $\left|x_{1_j}+ip_{1_j}\right\rangle$ with $j=1,2,\ldots,r_{\mathrm{min}}$. Using transmit-side beamforming, the $r_{\mathrm{min}}$ coherent states are transmitted through the atmospheric channel from the user 1’s $N_{t_1}$ transmit antennas to  user 2’s $N_{r_2}$ receive antennas. $r_k$ is the rank of the channel matrix between the $k$-th user and next participant; the detailed derivations and explanations of the channel matrix are described in Appendix \ref{appendixa}. The $x_{k_j}$ and $p_{k_j}$ with $k=1,2,\ldots,n$ represent the position and momentum quadrature of the coherent states. The variance of the modulated thermal noise is $V_S=1+\frac{2}{\exp{(hf/k_BT_e)}-1}$ and total quadrature variance $V=V_S+V_0$ \cite{39,40}, {where $V_0$ is the variance of Gaussian modulation}, {$T_e$} is the temperature, the $k_B=1.38\times 10^{-23}$J/K  and $h=6.626\times 10^{-34}$J·s are the  Boltzmann and Planck constants, respectively. 

\item \emph{Step 2:} User 2 employs receive-side beamforming with $N_{r_2}$ antennas to capture the coherent states sent from user 1. Once the transmit-receive beamforming is completed, the MIMO channel between the transmitter and receiver is decomposed into multiple parallel SISO subchannels.
 At the same time, user 2 generates $r_{\mathrm{min}}$ new thermal states with user 2's own THz source, which are subsequently modulated through AM and PM to generate $r_{\mathrm{min}}$ coherent states, denoted as  $\left|x_{2_j}+ip_{2_j}\right\rangle$. These locally prepared coherent states are then coupled with the received signal states using a highly asymmetric beam splitter (HABS) \cite{37,41}, ensuring that both sets of states occupy the same spatiotemporal mode.  The mixed states are then transmitted to the user 3. {For simplicity, all HABSs are assumed to be ideal and introduce no additional propagation loss, excess noise, or mode mismatch \cite{38}.}
 
\item \emph{Step 3:} Each subsequent user repeats the procedure described in step 2. Ultimately, the $r_{\mathrm{min}}$ mixed quantum states are transmitted to the dealer’s receive antennas. The $r_{\mathrm{min}}$ mixed states are $\left|\sum_{k=1}^{n}\sqrt{T_{k_j}}x_{k_j}+i\sum_{k=1}^{n}\sqrt{T_{k_j}}p_{k_j}\right\rangle$, where  $T_{k_j}$ denotes  the equivalent cumulative transmittance of the $j$-th SISO eigenchannel from user $k$ to the dealer through the subsequent forwarding path. Here, $T_{k_j}=1$ means there is perfect connection and $T_{k_j}=0$ means there is no connection. {The dealer then performs either homodyne or heterodyne detection on the $r_{\mathrm{min}}$ received mixed states to extract the quadrature information and generate the corresponding raw data. In the homodyne detection scheme, only one quadrature is measured, and the corresponding measurement outcome is denoted by $x_{d_j}$. In contrast, the heterodyne detection scheme simultaneously measures both the amplitude and phase quadratures, yielding the measurement outcomes denoted by $[x_{d_j},p_{d_j}]$.}
\item \emph{Step 4:} Steps 1-3 are repeated multiple times to gather sufficient raw data for parameter estimation and post-processing.
\end{itemize}

\noindent $Classical$ $post-processing$
\begin{itemize}
\item \emph{Step 5:} The dealer and all users reveal subsets of the raw data to estimate the respective channel transmittance $T_{k_j}$. These disclosed samples are discarded after parameter estimation.
\item \emph{Step 6:}  The dealer first assumes that user 1 is honest and treats the remaining $n-1$ users as untrusted. Another fresh subset of raw data is selected, and all users except user 1 are required to disclose their corresponding raw data. {Based on the revealed data and according to the adopted measurement strategy, the dealer performs a displacement operation on his measurement outcomes. In the heterodyne detection scheme, both quadratures are displaced as
	$x_{b_{1_j}}=x_{d_j}-\sum_{s\neq 1}^{n}\sqrt{T_{s_j}}x_{s_j}\mathrm{~and~}p_{b_{1_j}}=p_{d_j}-\sum_{s\neq 1}^{n}\sqrt{T_{s_j}}p_{s_j}$,
	whereas in the homodyne-detection scheme, only the measured quadrature is displaced. For instance, if the amplitude quadrature is measured, the displaced variable is given by
	$x_{b_{1_j}}=x_{d_j}-\sum_{s\neq 1}^{n}\sqrt{T_{s_j}}x_{s_j}$.} With the displaced measurement results and the disclosed data, {the resulting input--output relation can be modeled as an equivalent point-to-point CV-QKD between the dealer and user~1 for  SKR evaluation}.  If $R_{\mathrm{SISO}_{1_j}}>0$ for $1\leq j\leq r_{\mathrm{min}}$, the dealer and user 1 then compute the corresponding SKR with $\sum_{j=1}^{r_{\mathrm{min}}}R_{\mathrm{SISO}_{1_j}}$ using reverse reconciliation. Here, the SKR refers to the net number of secret bits that can be generated per channel use between user~1 and the dealer.
\item \emph{Step 7:} Steps 5 and 6 are repeated multiple times, during which the dealer sequentially chooses one user as the trusted party while treating all others as untrusted. The untrusted users are then required to disclose their measurement results. { Based on the displaced measurement outcomes and the disclosed data, the resulting input--output relation can be mapped to an equivalent CV-QKD model between the dealer and the trusted user, which is subsequently used for SKR analysis.} The SKR for each trusted user $k$ is computed as $\sum_{j=1}^{r_{\mathrm{min}}}R_{\mathrm{SISO}_{k_j}}$. {After completing this process for all users, the dealer obtains the SKR associated with each authorized user based on the corresponding equivalent CV-QKD model. }The system SKR of the CV-QSS protocol is then determined by the minimum achievable SKR among all users, i.e., $R_\mathrm{MIMO}=\min\{\sum_{j=1}^{r_{\mathrm{min}}}R_{\mathrm{SISO}_{1_j}},\sum_{j=1}^{r_{\mathrm{min}}}R_{\mathrm{SISO}_{2_j}},\ldots,\sum_{j=1}^{r_{\mathrm{min}}}R_{\mathrm{SISO}_{n_j}}\}$. {This metric quantifies the lower bound on the SKR achievable between the dealer and any authorized user in the QSS system.
}
\item \emph{Step 8:} The dealer applies the unused raw data to generate secret keys. After completing the standard post-processing procedures in CV-QKD, including error correction and privacy amplification \cite{42}, the dealer and each user obtain $r_{\mathrm{min}}$ independent secret-key shares ${K_{\mathrm{SISO}_{k_j}}}$.
\item \emph{Step 9:} The $r_{\mathrm{min}}$ messages $M_j$ are encrypted as $E_j = M_j \oplus K_j$ by the dealer, where $K_j = \bigoplus_{k=1}^n K_{\mathrm{SISO}_{k_j}}$ and $\bigoplus$ denotes the XOR operation. The dealer then broadcasts the ciphertext over the classical channel, which can be received by all participants in the system. Since each user only holds their own set of $r_{\mathrm{min}}$ keys $K_{\mathrm{SISO}_{k_j}}$,  the compromise of {secret-key shares} from a subset of users does not enable an eavesdropper to decrypt the ciphertext $E_j$.  As a result, the ciphertext cannot be decrypted unless all required secret-key shares are available.

\item \emph{Step 10:} When the users receive the ciphertext, they transmit their secret-key shares $K_{\mathrm{SISO}_{k_j}}$ to Charlie through authenticated secure classical channels for classical reconstruction.

\item \emph{Step 11:} Charlie uses the collected keys to decrypt the ciphertext by $M_j=E_j \bigoplus_{k=1}^n K_{\mathrm{SISO}_{k_j}}$ and publishes the recovered message $M_j$ to all users. This process is repeated $r_{\mathrm{min}}$ times, with Charlie decrypting one ciphertext segment during each iteration. {This process enables the recovered message to be shared among all authorized users after successful reconstruction.}

\end{itemize}
\begin{figure}[!t]
	\centering
	\includegraphics[width=\linewidth]{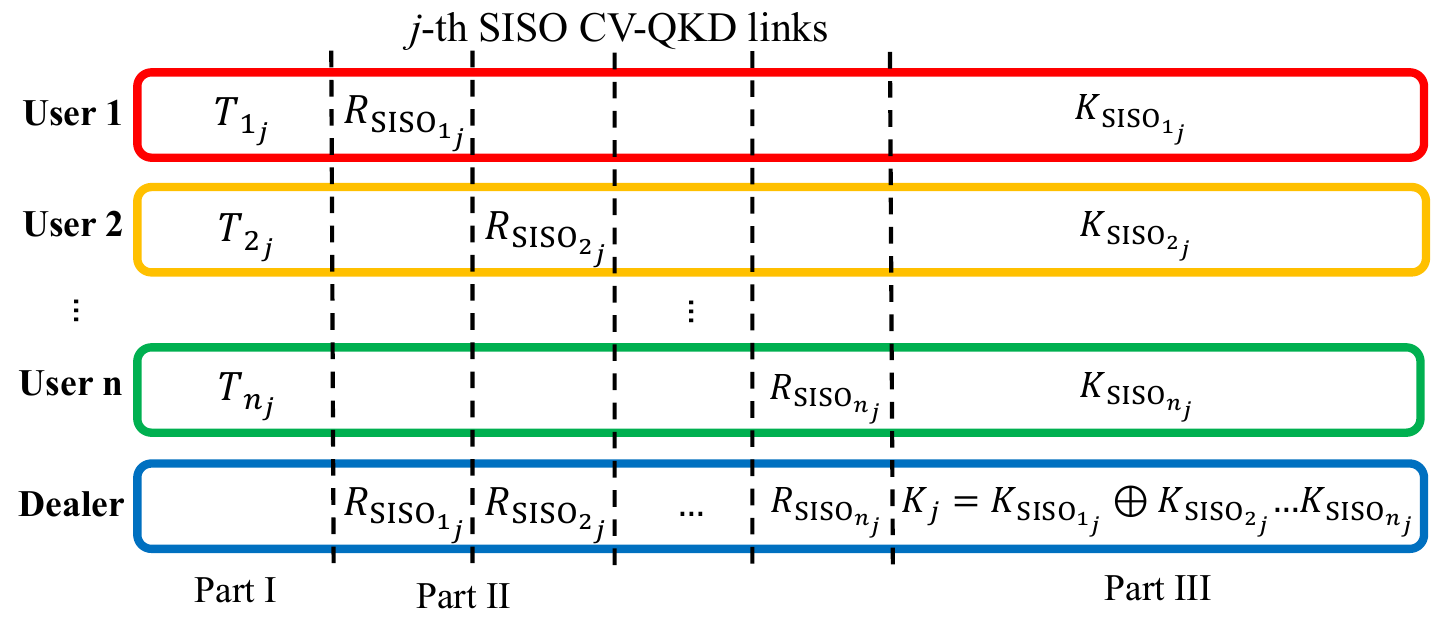}
	\caption{The usage of raw data generated by $n$ users and dealer in the $j$-th SISO channel.}
	\label{fig2}
\end{figure}

Figure \ref{fig2} shows how raw data from each SISO subchannel are partitioned into parameter estimation, SKR calculation, and final key generation. The raw data are divided into three segments: Part I is used to estimate the channel transmittance $T_{k_j}$, Part II is used to compute $R_{\mathrm{MIMO}}$, and Part III is used for final key generation. The raw data of Part I and Part II are randomly selected by both the dealer and the user. After completing step 7, the raw data of Part I and Part II are discarded.

{
It is worth emphasizing that the proposed scheme is not simply implemented by running independent point-to-point CV-QKD links followed by classical XOR operations. In a conventional XOR-over-QKD approach, the dealer must establish separate QKD links with each of the $n$ users through isolated or orthogonally multiplexed quantum channels, such that the quantum states associated with different users remain completely independent and exhibit no physical interaction or coherence. In contrast, the proposed QSS protocol relies on a sequential quantum transmission process.  As illustrated in Figure~\ref{fig1}, a spatio-temporal mode originates from user~1 and sequentially traverses the secure stations of all subsequent users. Each participant injects a locally prepared GMCS into the same propagating mode through a HABS. Consequently, the state received by the dealer contains a coherent superposition of all users' quadrature displacements, $\left| \sum_{k=1}^{n}\sqrt{T_{k_j}}x_{k_j} + i\sum_{k=1}^{n}\sqrt{T_{k_j}}p_{k_j}
\right\rangle $.
Thus, the users' information is physically combined within a single quantum mode before classical post-processing. The XOR operation is only used for the final classical reconstruction and does not replace the quantum multi-user sharing process.
}

\subsection{PMCS scheme}
\begin{figure}[!t]
	\centering
	\includegraphics[width=\linewidth]{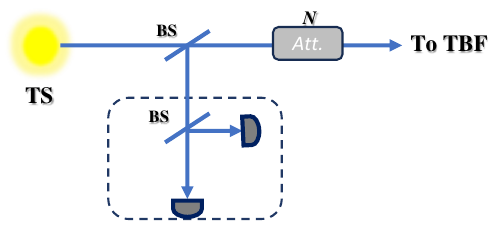}
	\caption{The generation process of PMCS. TS: terahertz source, BS: beam splitter, $Att.$: attenuator, $N$: the noise generated by $Att$.}
	\label{fig3}
\end{figure}
In practical GMCS schemes, the complex modulation requirements and stringent error tolerances demand modulators with high extinction ratios and exceptional stability. This significantly increases system cost and hinders experimental feasibility. To address this limitation, we further introduce the PMCS approach  \cite{38}. In this scheme, information is encoded with an attenuator and detector, avoiding the direct use of AM or PM. The PMCS protocol follows a procedure similar to that of the GMCS protocol, with the key distinction at the source (see Figure \ref{fig3}). Specifically, the THz source is first employed for user 1 to generate $r_{\mathrm{min}}$ thermal modes, $\text{mod}^j_{1_1}$ and $j=1,2,\ldots,r_{\mathrm{min}}$, which are injected into a beam splitter (BS) to produce two $r_{\mathrm{min}}$ spatial modes, $\text{mod}^j_{1_2}$ and $\text{mod}^j_{1_3}$. $\text{mod}^j_{1_3}$ is measured with heterodyne detection to obtain the $x$- and $p$-quadratures, whereas $\text{mod}^j_{1_2}$ are attenuated by an attenuator with noise $N^j_{1}$. The attenuated mode is then transmitted from user 1 to the next user via transmit antennas. At each subsequent stage, the receiving user couples their local mode $\text{mod}^j_{k_2}$ with the incoming signal into the same spatiotemporal mode and forwards it further. When the final mixed state arrives at the dealer, the channel transmittance will be estimated and the corresponding equivalent CV-QKD models between the dealer and the users are constructed for SKR evaluation. After reconciliation and post-processing, the corresponding secret-key shares are obtained between the dealer and the authorized users. This method reduces the state-preparation complexity and provides an implementation-oriented alternative to active Gaussian modulation within the adopted theoretical model.

\subsection{Relation to the previous works}
{Our work generalizes several previous studies} \cite{9,33,37,38}. When both the numbers of transmit and receive antennas are set to one and conventional optical sources and fiber links are considered, our GMCS and PMCS models reduce to those in \cite{37} and \cite{38}, respectively. Moreover, when the number of system users equals one, our GMCS model reduces to the conventional point-to-point QKD model \cite{9}, in which Alice transmits coherent states to Bob over an insecure quantum channel. Furthermore, we provide an explicit decryption method for QSS, whereas existing QSS protocols \cite{33,34,35,36,37,38} focus primarily on the encryption phase.

Our proposed QSS scheme leverages transmit–receive beamforming to decompose the MIMO channel into multiple parallel SISO channels. This structure enables the simultaneous encryption and transmission of $r_{\mathrm{min}}$ message segments, which does not exist in previous QSS protocols. Within this framework, steps 1-7 are particularly important because they ensure the secure and orderly execution of the protocol. If the estimated system SKR is non-positive, no secret key is extracted from the corresponding data block, and the key-generation process is aborted for that block. The derivation of the SKR, together with the theoretical assumptions  will be discussed in detail in the next section. 

\section{THEORETICAL ANALYSIS}

{
	
Analyzing the SKR $R_{\mathrm{MIMO}}$ is essential for evaluating both the security and performance of the proposed CV-QSS protocol. In the considered $(n,n)$-threshold CV-QSS scheme, the dealer is assumed to be a trusted party responsible for distributing secret shares to all legitimate users. Since reconstruction of the final secret requires the participation of all authorized users, the successful execution of the protocol depends on the availability of secure key generation between the dealer and each user.
In general, the system SKR is determined by the user with the lowest achievable SKR, $R_\mathrm{MIMO}=\min\{\sum_{j=1}^{r_{\mathrm{min}}}R_{\mathrm{SISO}_{1_j}},\sum_{j=1}^{r_{\mathrm{min}}}R_{\mathrm{SISO}_{2_j}},\ldots,\sum_{j=1}^{r_{\mathrm{min}}}R_{\mathrm{SISO}_{n_j}}\}$.

Under the adopted uniform deployment model, user~1 is located farthest from the dealer and its associated quantum signal experiences the largest number of intermediate forwarding operations. Therefore, within this idealized channel model, user~1 is regarded as the bottleneck user and provides a conservative benchmark for evaluating the system performance. Accordingly, the following theoretical analysis focuses on the equivalent channel between user~1 and the dealer. This reduction is used for analytical tractability and SKR evaluation under the considered deployment assumptions.
}

{ 
It should be noted that although theoretical analyses commonly assume that the farthest link yields the lowest key rate, real-world deployments may not always conform to this assumption. Factors such as carrier frequency response, hardware imperfections, and multipath propagation can significantly influence the SKR. Therefore, in experimental evaluations, SKRs should be assessed on a link-by-link basis.  In this work, we adopt a modeling approach similar to that in Refs. \cite{9}, where perfect beamforming is assumed, while practical engineering factors in ultra-massive MIMO arrays, such as channel fluctuations, synchronization, phase noise suppression, and thermal management, are not explicitly considered.  }
\begin{figure}[!t]
	\centering
	\includegraphics[width=\linewidth]{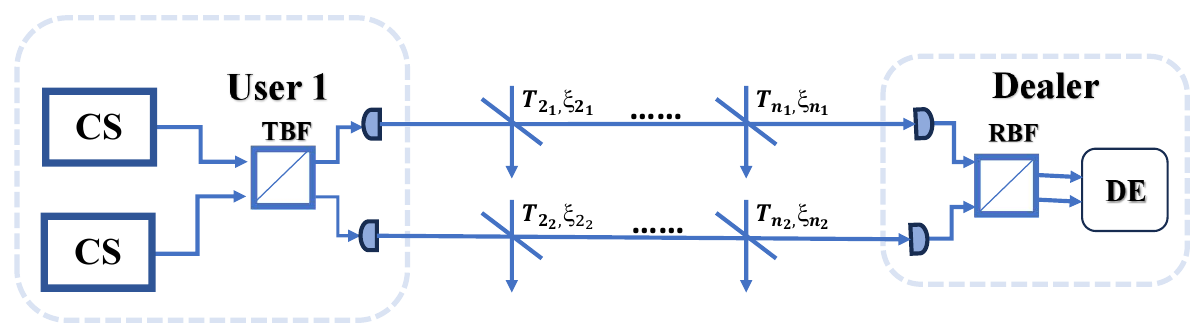}
	\caption{Equivalent model for computing the SKR of MIMO THz QSS protocol. CS: coherent state, TBF and RBF: transmit and receive beamforming, DE: detection.}
	\label{fig10}
\end{figure}

 The equivalent model is shown in Figure \ref{fig10}, where user 1 employs transmit-side beamforming to send modulated quantum states to the dealer, who performs receive-side beamforming to collect quantum signals. {
 	 Meanwhile, we consider a worst-case scenario in which the remaining $n-1$ untrusted users are allowed to fully collude with one another. The impact of collusion attacks is incorporated into the security analysis as additional excess noise, denoted by $\xi_{k_j}$, which is modeled as \cite{37}
\begin{equation}\label{188}
	\xi_{k_j}=\frac{T_{k_j}}{T_{1_j}}\xi_{1_j}.\end{equation}
The total channel-equivalent excess noise $\hat{\xi}_{1_j}$ in the $j$-th SISO subchannel between the first user and the dealer can be modeled as the accumulation of noise contributions from all intermediate links. For GMCS protocol, total excess noise $\hat{\xi}_{1_j}$ is given by
\begin{equation}\label{29}
	\hat{\xi}_{1_j}^G=\sum_{k=1}^n\xi_{k_j}.\end{equation}
For PMCS protocol, it is given by
\begin{equation}\label{30}
	\hat{\xi}_{1_j}^P=\sum_{k=1}^n(\xi_{k_j}+N_{k_j}).\end{equation}

 Increased excess noise reduces the correlation between the legitimate parties and consequently lowers the achievable SKR. Therefore, the resulting SKR provides a conservative performance estimate under the adopted collusion attack model.} The equivalent model can be given by
\begin{equation}\mathbf{H_{D,1}}=\mathbf{H_n}\mathbf{F_{ n}}\mathbf{H_{n-1}}\mathbf{F_{ n-1}}\cdots\mathbf{F_{ 2}}\mathbf{H_{1}}\end{equation}
where $\mathbf{F}_{k}$ denotes the effective mode-selection matrix induced by the forwarding operation at user~$k$. Under the ideal HABS assumption, it maps the received $N_{r_k}$ spatial modes to the transmitted $N_{t_k}$ spatial modes while preserving the first $v_a=\min\{N_{t_k},N_{r_k}\}$ available modes. {Under the perfect-HABS assumption, it can be written as}
\begin{equation}	\mathbf{F}_{k}=
	\begin{bmatrix}
		\boldsymbol{I}_{v_a} & \mathbf{0}_{v_a \times (N_{r_k}-v_a)} \\
		\mathbf{0}_{(N_{t_k}-v_a)\times v_a} & \mathbf{0}_{(N_{t_k}-v_a)\times (N_{r_k}-v_a)}
\end{bmatrix},\end{equation}
where $v_a= \mathrm{min}\{N_{t_k},N_{r_k}\}$. The overall matrix $\mathbf{H_{D,1}}$ can be characterized via singular-value decomposition as proposed by \cite{9999}
 \begin{equation}\label{1}
 	\mathbf{H_{D,1}}=\mathbf{U_D}\mathbf{\Sigma_{r_{\mathrm{min}}}}\mathbf{V_1} ^{\dagger},\end{equation}
 where $\mathbf{U_D}\in\mathbb{C}^{N_{r_D}\times N_{r_D}}$ and $\mathbf{V_1}\in\mathbb{C}^{N_{t_1}\times N_{t_1}}$ are unitary transformation matrices. {Unitary transformations do not introduce additional noise and preserve the canonical commutation relations of quantum states. Physically, such transformations can be implemented using an optical device composed of beam splitters and phase shifters \cite{9998}.} The matrix $\mathbf{\Sigma_{r_{\mathrm{min}}}}$  is
 \begin{equation}
 	\small
 	\mathbf{\Sigma_{r_{\mathrm{min}}}}=
 	\begin{bmatrix}
 		\operatorname{diag}\left\{\sqrt{T_{1_1}},\ldots,\sqrt{T_{1_{r_{\mathrm{min}}}}}\right\} & \mathbf{0}_{r_{\mathrm{min}}\times(N_{t_1}-r_{\mathrm{min}})} \\
 		\mathbf{0}_{(N_{r_D}-r_{\mathrm{min}})\times r_{\mathrm{min}}} & \mathbf{0}_{(N_{r_D}-r_{\mathrm{min}})\times(N_{t_1}-r_{\mathrm{min}})}
 \end{bmatrix},\end{equation}
 where $\sqrt{T_{1_j}}$ is the $j$-th nonzero singular value of $\mathbf{H}_{D,1}$.
 
 At the transmitter side, user 1 applies transmit-side beamforming with $\mathbf{V_1}$, while at the receiver side, the dealer performs receive-side beamforming with $\mathbf{U_D}^{\dagger}$. The relationship between the input mode of the user 1 and the output mode of the dealer in a MIMO system can be expressed as follows
 \begin{equation}\label{3}
 	\mathbf{mod_{D}^{out}}=\mathbf{U_D}^\dagger \mathbf{H_{D,1}}\mathbf{V_1}\mathbf{mod_{1}^{in}}+\mathbf{U_D}^\dagger \mathbf{U_D}\mathbf{S_1}\mathbf{mod_{E}}.\end{equation}
By substituting equation (\ref{1}) into equation (\ref{3}) and $\mathbf{U_D}^{\dagger}\mathbf{U_D}$ and $\mathbf{V_1}^{\dagger}\mathbf{V_1}$ are $N_{r_D}\times N_{r_D}$ and $N_{t_1}\times N_{t_1}$ identity matrices, respectively, we have
\begin{equation}\label{4}
	\mathbf{mod_{D}^{out}}=\mathbf{\Sigma_{r_{\mathrm{min}}}}\mathbf{mod_{1}^{in}}+\mathbf{S_1}\mathbf{mod_{E}},\end{equation}
where $\mathbf{mod_{D}^{out}}=[\text{mod}_{D}^{1,\text{out}},\ldots,\text{mod}_{D}^{N_{r_D},\text{out}}]^{T}$ and $\mathbf{mod_{1}^{in}}=[\text{mod}_{1}^{1,\text{in}},\ldots,\text{mod}_{1}^{r_{\mathrm{min}},\text{in}},\mathbf{0}_{1 \times (N_{t_1}-r_{\mathrm{min}})}]^{T},$ denote the dealer’s receive modes and the first user’s transmit modes, respectively. {$\mathbf{mod_{E}}=[\text{mod}_{E}^{1},\ldots,\text{mod}_{E}^{N_{r_D}}]^{T}$} denotes the equivalent environmental ancillary-mode vector associated with the decomposed bosonic channel. In the adopted worst-case collusion model, these environmental modes are assumed to be accessible to the enlarged adversarial system, which includes the external eavesdropper Eve and the colluding untrusted users. {$\mathbf{S}_1= \operatorname{diag}\{ \sqrt{1-T_{1_1}},\ldots,\sqrt{1-T_{1_{r_{\mathrm{min}}}}}, \underbrace{1,\ldots,1}_{N_{r_D}-r_{\mathrm{min}}} \}$} is a diagonal matrix. By substituting the corresponding parameters into equation (\ref{4}), we obtain

\begin{equation}
\small
\begin{bmatrix}\text{mod}_{D}^{1,\text{out}}\\ \vdots\\\text{mod}_{D}^{r_{\mathrm{min}},\text{out}} \\ \text{mod}_{D}^{r_{\mathrm{min}}+1,\text{out}} \\ \vdots\\ \text{mod}_{D}^{N_{r_D},\text{out}} \end{bmatrix}=\begin{bmatrix}\sqrt{T_{1_1}}\text{mod}_{1}^{1,\text{in}}+\sqrt{1-T_{1_1}}\text{mod}_{E}^{1}\\ \vdots\\\sqrt{T_{1_{r_{\mathrm{min}}}}}\text{mod}_{1}^{r_{\mathrm{min}},\text{in}}+\sqrt{1-T_{1_{r_{\mathrm{min}}}}}\text{mod}_{E}^{r_{\mathrm{min}}}\\ \text{mod}_{E}^{r_{\mathrm{min}}+1} \\ \vdots\\ \text{mod}_{E}^{N_{r_D}} \end{bmatrix},
\end{equation}
{where  $\mathrm{mod}_{D}^{j,\mathrm{out}}$, $\mathrm{mod}_{1}^{j,\mathrm{in}}$, and $\mathrm{mod}_{E}^{j}$ denote the annihilation operators corresponding to the dealer's output mode, the first user's input mode, and the equivalent ancillary environmental mode. In the adopted worst-case joint attack model, this enlarged adversarial system includes both the external eavesdropper Eve and the colluding $(n-1)$ untrusted users.  The notation $(\cdot)^\dagger$ denotes the Hermitian conjugate operation. Accordingly, $\left(\mathrm{mod}_{D}^{j,\mathrm{out}}\right)^\dagger$, $\left(\mathrm{mod}_{1}^{j,\mathrm{in}}\right)^\dagger$, and $\left(\mathrm{mod}_{E}^{j}\right)^\dagger$ represent the corresponding creation operators. }

{For the $j$-th output mode and $1 \leq j \leq r_{\mathrm{min}}$, each output depends only on its corresponding single input mode and remains independent of other input modes. Moreover, $\text{mod}_{D}^{j,\text{out}} =\sqrt{T_{1_j}}\text{mod}_{1}^{j,\text{in}}+\sqrt{1-T_{1_j}}\text{mod}_{E}^{j}$ corresponds to the standard  input–output relation of SISO channel \cite{10000}.  Therefore, the MIMO channel between user 1 and the dealer can be decomposed into $r_{\mathrm{min}}$ parallel SISO channels.}

 {To further verify that these decomposed SISO subchannels are physically valid bosonic channels in the CV quantum regime, we examine the preservation of the canonical commutation relations. For $1\leq m,n\leq r_{\mathrm{min}}$,
	  we have
	   \begin{equation}\label{111} \begin{aligned} & \left[ \mathrm{mod}_{D}^{m,\mathrm{out}}, \left( \mathrm{mod}_{D}^{n,\mathrm{out}} \right)^\dagger \right] \\ =& \left[ \sqrt{T_{1_m}} \mathrm{mod}_{1}^{m,\mathrm{in}} + \sqrt{1-T_{1_m}} \mathrm{mod}_{E}^{m}, \right. \\ & \left. \sqrt{T_{1_n}} \left( \mathrm{mod}_{1}^{n,\mathrm{in}} \right)^\dagger + \sqrt{1-T_{1_n}} \left( \mathrm{mod}_{E}^{n} \right)^\dagger \right] \\ =& \sqrt{T_{1_m}T_{1_n}} \left[ \mathrm{mod}_{1}^{m,\mathrm{in}}, \left( \mathrm{mod}_{1}^{n,\mathrm{in}} \right)^\dagger \right] \\ & + \sqrt{T_{1_m}(1-T_{1_n})} \left[ \mathrm{mod}_{1}^{m,\mathrm{in}}, \left( \mathrm{mod}_{E}^{n} \right)^\dagger \right] \\ & + \sqrt{(1-T_{1_m})T_{1_n}} \left[ \mathrm{mod}_{E}^{m}, \left( \mathrm{mod}_{1}^{n,\mathrm{in}} \right)^\dagger \right] \\ & + \sqrt{(1-T_{1_m})(1-T_{1_n})} \left[ \mathrm{mod}_{E}^{m}, \left( \mathrm{mod}_{E}^{n} \right)^\dagger \right]. \end{aligned}  \end{equation}
	    Since the input modes and  ancillary modes are independent, the cross-commutators vanish, namely, \cite{10000} \begin{equation}\label{112} \left[ \mathrm{mod}_{1}^{m,\mathrm{in}}, \left( \mathrm{mod}_{E}^{n} \right)^\dagger \right] = \left[ \mathrm{mod}_{E}^{m}, \left( \mathrm{mod}_{1}^{n,\mathrm{in}} \right)^\dagger \right] =0. \end{equation}
	     Moreover, both the input modes and  ancillary modes satisfy the canonical bosonic commutation relations, \begin{equation}\label{113} \left[ \mathrm{mod}_{1}^{m,\mathrm{in}}, \left( \mathrm{mod}_{1}^{n,\mathrm{in}} \right)^\dagger \right] = \left[ \mathrm{mod}_{E}^{m}, \left( \mathrm{mod}_{E}^{n} \right)^\dagger \right] = \delta_{mn}, \end{equation}
	      where $\delta_{mn}$ denotes the Kronecker delta. Substituting equations (\ref{112}) and (\ref{113}) into equation (\ref{111}), we obtain
	       \begin{equation} \begin{aligned} & \left[ \mathrm{mod}_{D}^{m,\mathrm{out}}, \left( \mathrm{mod}_{D}^{n,\mathrm{out}} \right)^\dagger \right] \\ =& \left( \sqrt{T_{1_m}T_{1_n}} + \sqrt{(1-T_{1_m})(1-T_{1_n})} \right) \delta_{mn}. \end{aligned} \end{equation}
	        Since $\delta_{mn}=0$ for $m\neq n$ and $\delta_{mn}=1$ for $m=n$, it follows that \begin{equation} \sqrt{T_{1_m}T_{1_n}}\delta_{mn} = T_{1_m}\delta_{mn}, \end{equation}
	         and 
	         \begin{equation} \sqrt{(1-T_{1_m})(1-T_{1_n})}\delta_{mn} = (1-T_{1_m})\delta_{mn}. \end{equation}
	          Hence, 
	          \begin{equation} \begin{aligned} \left[ \mathrm{mod}_{D}^{m,\mathrm{out}}, \left( \mathrm{mod}_{D}^{n,\mathrm{out}} \right)^\dagger \right] &= \left( T_{1_m} + 1-T_{1_m} \right) \delta_{mn} \\ &= \delta_{mn}. \end{aligned} \label{CCR} \end{equation} 
	          Therefore, the canonical commutation relations are preserved after the channel decomposition, confirming that the resulting SISO subchannels are physically valid bosonic channels in the CV quantum regime.}

{Furthermore, among various eavesdropping strategies, including individual attacks, coherent attacks, and Gaussian collective attacks, Gaussian collective attacks have been proven to be optimal for CV quantum communication systems \cite{27}. Accordingly, throughout this paper, the security performance of the proposed scheme is evaluated under Gaussian collective attacks performed by an enlarged adversarial system, which consists of the external eavesdropper Eve and the colluding untrusted users.}

To establish a conservative performance benchmark, the lower bound of the SKR for the MIMO CV-QSS protocol can be estimated by analyzing the equivalent CV-QKD model between user~1 and the dealer. We first derive the asymptotic SKR, which serves as the theoretical upper bound on the protocol’s performance. According to step~7, this asymptotic SKR (in bit/use) is given by:
\begin{equation}\label{6}
	R_{\mathrm{MIMO}}^\mathrm{A}=\sum_{j=1}^{r_{\mathrm{min}}}R_{\mathrm{SISO}_{1_j}}^\mathrm{A},\end{equation}
where $R_{\mathrm{SISO}_{1_j}}^\mathrm{A} = \beta I (U_{1_j}:D_{1_j}) - \chi(D_{1_j}:E)$ is the SKR of the $j$-th SISO channel,  $\beta$ is the reverse reconciliation efficiency, and $r_{\mathrm{min}}$ denotes the number of equivalent SISO channels established.  {Let the transmission distance between user~1 and the dealer be $d_{U_1D}$, and assume that the $n$ users are uniformly distributed with equal spacing.} The distance between the $k$-th user and the dealer can be expressed as
\begin{equation}d_{U_kD}=\frac{n-k+1}{n}d_{U_1D}.\end{equation}
 The transmittance of the $j$-th SISO channel between the $k$-th user and dealer can be expressed as \cite{44}
\begin{equation}T_{k_j}=
	\begin{cases}
		(\frac{\lambda}{4\pi d_{U_kD}})^{2}10^{-\frac{\delta d_{U_kD}}{10}}G_{t_{k}}G_{r_{D}}, & j=1, \\
		F(\frac{\lambda}{4\pi d_{U_kD}})^{2}10^{-\frac{\delta d_{U_kD}}{10}}G_{t_{k}}G_{r_{D}}, & j=2,3,\ldots,r_{\mathrm{min}},
	\end{cases}\end{equation}
where $G_{t_k}=N_{t_k}G_a,G_{r_D}=N_{r_D}G_a$ are the uniform linear array gains for $k$-th user and dealer, and $G_a$ is the gain of each transmit or receive antenna, $F$ is total attenuation of the non-line-of-sight (NLoS) path, $\delta$ represents the atmospheric absorption loss (dB/km) and $\lambda$ denotes the wavelength, whose product with the carrier frequency equals the speed of light. We set $j=1$ as the line-of-sight (LoS) path, while the others are NLoS paths. To evaluate the mutual information $I (U_{1_j}:D_{1_j})$  and the Holevo bound $\chi(D_{1_j}:E)$, we introduce the equivalent input-referred noise variance $W_{1_j}$ of the channel. For GMCS protocol, $W_{1_j}$ is given by \cite{45}
\begin{equation} \label{12}
	W_{1_j}^G=\frac{T_{1_j}(\hat{\xi}_{1_j}^G-1)+1}{1-T_{1_j}}.
\end{equation}
For PMCS protocol, $W_{1_j}$ is given by
\begin{equation} \label{13}
	W_{1_j}^P=\frac{T_{1_j}(\hat{\xi}_{1_j}^P-1)+1}{1-T_{1_j}}.
\end{equation}
where $\hat{\xi}_{1_j}^G$ and $\hat{\xi}_{1_j}^P$ are given in equations (\ref{29}) and (\ref{30}). 
%

The  $I (U_{1_j}: D_{1_j}) $ is the Shannon mutual information between user 1 and dealer in the $j$-th SISO link. For homodyne detection, it is given by \cite{49}
\begin{equation}I (U_{1_j}:D_{1_j})=\frac{1}{2}\log_{2}\left[1+\frac{T_{1_j}V_{0}}{T_{1_j}V_S+(1-T_{1_j})W_{1_j}}\right].\end{equation}
For heterodyne detection, it is given by
\begin{equation}I (U_{1_j}:D_{1_j})=\log_{2}\left[1+\frac{T_{1_j}V_{0}}{T_{1_j}V_S+(1-T_{1_j})W_{1_j}+1}\right].\end{equation}

On the other hand, the Holevo bound $\chi(D_{1_j}:E)$, which represents the maximum information that the eavesdropper can obtain about the dealer's measurement outcome in the $j$-th SISO link, is \cite{47,48}
\begin{equation}\chi(D_{1_j}:E)=S_E-S_{E|D},\end{equation}
where $S_E$ is the von Neumann entropy of eavesdropper’s total state and $S_{E|D}$ is the von Neumann entropy of eavesdropper’s conditional state. { Under Gaussian collective attacks, it can be expressed as} {\cite{999,998}}
\begin{equation}\chi(D_{1_j}:E)=G\left(\lambda_1^j\right)+G\left(\lambda_2^j\right)-G\left(\lambda_3^j\right)-G\left(\lambda_4^j\right),\end{equation}
{where  $G\left(*\right)=\left(\frac{*+1}{2}\right)\log_2\left(\frac{*+1}{2}\right)-\left(\frac{*-1}{2}\right)\log_2\left(\frac{*-1}{2}\right),$
the $\lambda_{1,2}^j$ are the symplectic eigenvalues of matrix $\Sigma_E^{j}$ and $\lambda_{3,4}^j$ are the symplectic eigenvalues of matrix $\Sigma_{E|D}^{j}$. The $\Sigma_E^{j}$ is independent of the specific measurement strategy employed at the receiver. It is given by  {\cite{997}}
\begin{equation}
	\begin{aligned}
		\Sigma_E^{j}
		&=
		\begin{pmatrix}
			\big[(1-T_{1_j})V+T_{1_j}W_{1_j}\big]\boldsymbol{I}_2
			&
			\sqrt{T_{1_j}(W_{1_j}^2-1)}\boldsymbol{Z}
			\\
			\sqrt{T_{1_j}(W_{1_j}^2-1)}\boldsymbol{Z}
			&
			W_{1_j}\boldsymbol{I}_2
		\end{pmatrix}
		\\
		&=
		\begin{pmatrix}
			a_{1_j}\boldsymbol{I}_2 & c_{1_j}\boldsymbol{Z} \\
			c_{1_j}\boldsymbol{Z} & b_{1_j}\boldsymbol{I}_2
		\end{pmatrix}.
	\end{aligned}
\end{equation}
Thus, $\lambda_{1,2}^j$ can be given by 
\begin{equation}\lambda_{1,2}^j=\frac{1}{2}\left(\varepsilon_{1_j}\pm[b_{1_j}-a_{1_j}]\right),\end{equation}
and 
\begin{equation}\label{99}
\varepsilon_{1_j}=\sqrt{(a_{1_j}+b_{1_j})^2-4c_{1_j}^2}.
\end{equation}
When the dealer performs a homodyne detection measurement, the $\Sigma_{E|D}^{j}$ is given by {\cite{45}}
\begin{equation}\label{98}
	\Sigma_{E|D}^{j,ho}=\Sigma_{E}^j-\frac{\Sigma_{C_{1,j}}\Pi\Sigma_{C_{1,j}}^{T}}{T_{1_j}V+(1-T_{1_j})W_{1_j}}.\end{equation}
When the dealer performs a heterodyne measurement,  the $\Sigma_{E|D}^{j}$ is given by {\cite{995}}
\begin{equation}\label{97}
	\Sigma_{E|D}^{j,he}=\Sigma_{E}^j-\frac{\Sigma_{C_{1,j}}\Sigma_{C_{1,j}}^{T}}{T_{1_j}V+(1-T_{1_j})W_{1_j}+1},\end{equation}
where 
\begin{equation} \label{96}
	\Sigma_{C_{1,j}}=
	\begin{pmatrix}
		\sqrt{T_{1_j}\!\left(1-T_{1_j}\right)}
		\left(W_{1_j}-V\right)\boldsymbol{I}_2
		\\
		\\
		\sqrt{1-T_{1_j}}\sqrt{W_{1_j}^{\,2}-1}\,\boldsymbol{Z}
	\end{pmatrix}
,\end{equation}
and
\begin{equation}\label{95}
	\mathbf{\Pi}=
	\begin{bmatrix}
		1 & 0 \\
		0 & 0
\end{bmatrix}.\end{equation}
Based on equations (\ref{98}), (\ref{96}) and (\ref{95}), we have
\begin{equation}
	\Sigma_{E|D}^{j,ho}=
	\begin{bmatrix}
		\mathbf{A}_j & \mathbf{C}_j \\
		\mathbf{C}_j^T & \mathbf{B}_j
	\end{bmatrix},
\end{equation}

with

\begin{equation}
	\mathbf{A_j}=
	\begin{bmatrix}
		\frac{V W_{1_j}}{T_{1_j}(V-W_{1_j})+W_{1_j}} & 0 \\
		\\
		0 & -T_{1_j}V+V+T_{1_j}W_{1_j}
	\end{bmatrix},
\end{equation}

\begin{equation}
	\mathbf{B_j}=
	\begin{bmatrix}
		\frac{V W_{1_j}T_{1_j}-T_{1_j}+1}{T_{1_j}V-T_{1_j}W_{1_j}+W_{1_j}} & & 0 \\
		\\
		0 & & W_{1_j}
	\end{bmatrix},
\end{equation}

\begin{equation}
	\mathbf{C_j}=
	\begin{bmatrix}
		\frac{\sqrt{T_{1_j}}V\sqrt{W_{1_j}^2-1}}{T_{1_j}(V-W_{1_j})+W_{1_j}} & & 0 \\
		\\
		0 & & -\sqrt{T_{1_j}}\sqrt{W_{1_j}^2-1}
	\end{bmatrix}.
\end{equation}

For homodyne detection, the symplectic eigenvalues $\lambda_{3,4}^{j}$ can be given by
\begin{equation}
	\lambda_{3,4}^{j,ho}
	=
	\sqrt{
		\frac{1}{2}
		\left(
		\Theta_j
		\pm
		\sqrt{\Theta_j^2 - 4\kappa_j}
		\right)
	},
\end{equation}
where $\Theta_j=\det(\mathbf{A_j})+\det(\mathbf{B_j})+2\det(\mathbf{C_j})$ and $\kappa_j=\det\Sigma_{E|D}^{j,ho}$.
Based on equations. (\ref{97}) and (\ref{96}), we have
\begin{equation}
	\begin{aligned}
			\Sigma_{E|D}^{j,he}
		=
		\begin{pmatrix}
			d_{1_j}\boldsymbol{I}_2 & f_{1_j}\boldsymbol{Z} \\
			f_{1_j}\boldsymbol{Z} & e_{1_j}\boldsymbol{I}_2
		\end{pmatrix}.
	\end{aligned}
\end{equation}
where
\begin{equation}
d_{1_j}
=
\frac{
	-T_{1_j}V+W_{1_j}V+V+T_{1_j}W_{1_j}
}{
	T_{1_j}V-T_{1_j}W_{1_j}+W_{1_j}+1
},\end{equation}
\begin{equation}
e_{1_j}
=
\frac{V W_{1_j} T_{1_j}-T_{1_j}+W_{1_j}+1}
{T_{1_j}V-T_{1_j}W_{1_j}+W_{1_j}+1},
\end{equation}
\begin{equation}
f_{1_j}
=
\frac{
	\sqrt{T_{1_j}}(V+1)\sqrt{W_{1_j}^{2}-1}
}{
	T_{1_j}(V-W_{1_j})+W_{1_j}+1
}.
\end{equation}
For heterodyne detection, we get the symplectic eigenvalues $\lambda_{3,4}^{j}$ by
\begin{equation}\lambda_{3,4}^{j,he}=\frac{1}{2}\left(\tau_{1_j}\pm[e_{1_j}-d_{1_j}]\right),\end{equation}
and 
\begin{equation}
	\tau_{1_j}=\sqrt{(d_{1_j}+e_{1_j})^2-4f_{1_j}^2}.
\end{equation}
}
Next, we summarize the eight SKR expressions obtained from Propositions 1 and 2 for the proposed GMCS and PMCS schemes.

\begin{proposition} 
The asymptotic SKR expressions for the GMCS protocol under homodyne and heterodyne detection schemes are given by
\begin{equation}\label{20}
	\begin{aligned}R_{\mathrm{MIMO}}^\mathrm{A}=&\sum_{j=1}^{r_{\mathrm{min}}}\frac{\beta+\beta\Omega}{2}\log_{2}\left[1+\frac{T_{1_j}V_{0}}{T_{1_j}V_S+(1-T_{1_j})W_{1_j}^G+\Omega}\right]\\&-G\left(\lambda_1^j\right)-G\left(\lambda_2^j\right)+G\left(\lambda_3^{j,\mathrm{ho/he}}\right)+G\left(\lambda_4^{j,\mathrm{ho/he}}\right), \end{aligned} \end{equation}
where 
\begin{equation}
	\Omega=
	\begin{cases}
		0, & \text{homodyne detection} \\
		1, & \text{heterodyne detection}
	\end{cases}.
\end{equation}

\end{proposition}


In practical QSS systems, parameter estimation and privacy amplification must be taken into account, as statistical fluctuations arising from the exchange of a finite number of quantum signals can lead to discrepancies between the estimated channel parameters and their true values. Privacy amplification is applied to ensure that the final key is secure in an information-theoretic sense. It eliminates any residual information an eavesdropper may have gained through the quantum channel and classical error-correction process. 

According to Appendix \ref{appendixb}, for the {composable} finite-size case, we provide the following result:

\begin{proposition}
The  {composable finite-size SKR expressions} of GMCS under homodyne and heterodyne detection schemes are given by
\begin{equation}\label{22}
\begin{aligned}
		& R_{\mathrm{MIMO}}^{\mathrm{F}} \\
		=& \sum_{j=1}^{r_{\mathrm{min}}}\frac{N}{M}\Bigg[
		\frac{\beta+\beta\Omega}{2}
		\log_{2}\!\left(
		1+\frac{T_{1_j}V_{0}}{T_{1_j}V_S+(1-T_{1_j})W_{1_j}^G+\Omega}
		\right)  \\
		&\quad -G\!\left(\hat{\lambda}_1^j\right)
		-G\!\left(\hat{\lambda}_2^j\right)
		+G\!\left(\hat{\lambda}_3^{j,\mathrm{ho/he}}\right)+G\!\left(\hat{\lambda}_4^{j,\mathrm{ho/he}}\right)
		\Bigg] \\
		&-\frac{\sqrt{N}}{M}\Delta_{AEP}-\Delta_{PA}. 
\end{aligned}
\end{equation}
The values $\hat{\lambda}_1^j,\hat{\lambda}_2^j$, $\hat{\lambda}_3^j$ and $\hat{\lambda}_4^j$ are given in Appendix \ref{appendixb}.
\end{proposition}


Moreover, the SKR expressions for the PMCS-based MIMO scheme can be obtained directly from the GMCS results through simple parameter substitutions. Specifically, replacing $\hat{\xi}_{1_j}^{G}$ with  $\hat{\xi}_{1_j}^{P}$ yields the asymptotic SKR of PMCS, while replacing $T_{1_j,\mathrm{min}}^G$, $\hat{\xi}_{1_j,\mathrm{max}}^G$ with $T_{1_j,\mathrm{min}}^P$, $\hat{\xi}_{1_j ,\mathrm{max}}^P$ gives the corresponding { composable} finite-size SKR. The quantities $\hat{\xi}_{1_j}^{G}$ and $\hat{\xi}_{1_j}^{P}$ are defined in equations (\ref{29}) and (\ref{30}), respectively. The parameters $T_{1_j,\mathrm{min}}^{G}$ and $\hat{\xi}_{1_j,\mathrm{max}}^{G}$ are defined in equation (\ref{52}), whereas $T_{1_j,\mathrm{min}}^{P}$ and $\hat{\xi}_{1_j,\mathrm{max}}^{P}$ are defined in equation (\ref{53}).

%
%

\begin{figure}[t]
	\centering
	\includegraphics[width=\linewidth]{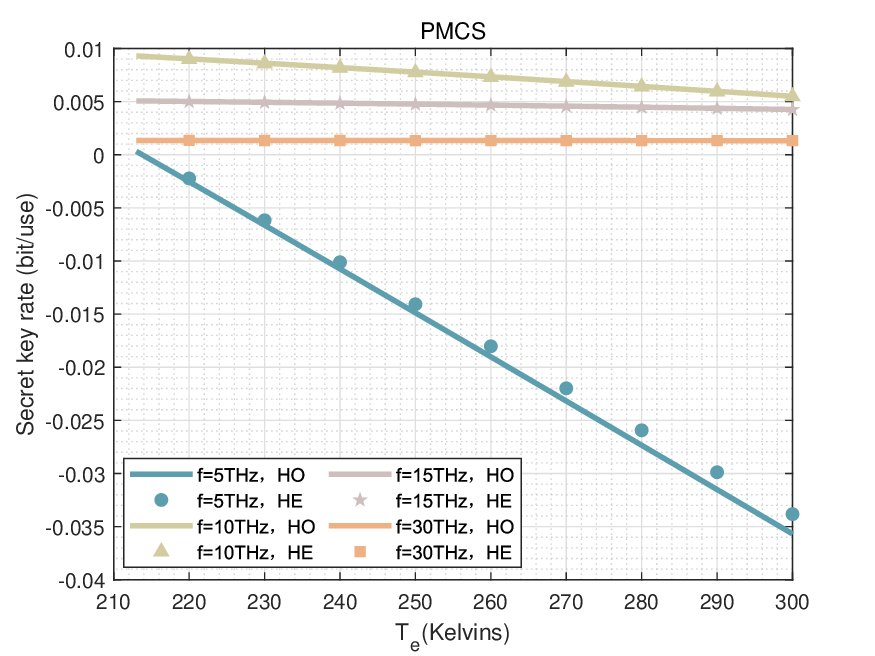}
	\caption{SKR of the PMCS protocol as a function of temperature at different carrier frequencies when the transmission distance approaches zero, where solid lines represent homodyne detection, and markers denote heterodyne detection. HO: homodyne, HE: heterodyne detection. $n=2$, $N_a=0.005$.}
	\label{fig4}
\end{figure}

\section{SIMULATION AND DISCUSSION}
In this section, we evaluate the proposed MIMO THz CV-QSS protocol in atmospheric channels through numerical simulations. For analytical convenience, we set $N_t = N_{t_1} = N_{t_2} = \cdots = N_{t_n}$, $N_r = N_{r_2} = \cdots = N_{r_n} = N_{r_D}$, excess noise $\xi = \xi_{1_1} = \xi_{1_2} = \cdots = \xi_{1_{r_{\mathrm{min}}}}$, and attenuator noise $N_a = N^1_{1} = N^2_{1} = \cdots = N^{r_{\mathrm{min}}}_{n}$. To maximize the multipath gain, we set $L=L_1 = L_2 =\cdots=L_n= \min(N_t, N_r)$ \cite{1011}. The system parameters are configured as follows: reverse reconciliation efficiency $\beta = 0.98$, modulation variance $V_0 = 100$,  excess noise $\xi=0.005$,  Boltzmann constant {$k_B=1.38\times 10^{-23}$ J/K}, Planck constant $h=6.626\times 10^{-34}$ J·s, amplitude attenuation $F=0.98$, antenna gain $G_a=30$ dBi and atmospheric absorption losses $\delta$ at carrier frequencies 1~THz$\leq f\leq 10$~THz and 10~THz$\textless f\leq 30$~THz, corresponding to 1000 dB/km and 50 dB/km respectively \cite{49}.  {It should be noted that the numerical results presented in this section are obtained under idealized assumptions and are intended to provide insight into the potential performance of the proposed scheme.}

Figure~\ref{fig4} illustrates the variation of the SKR with ambient temperature for the PMCS protocol at an approximately zero transmission distance. The results indicate that at $f = 5$~THz, the system fails to achieve a positive SKR under room-temperature conditions due to the dominance of thermal noise. In contrast, at higher carrier frequency (10, 15, and 30~THz), the system maintains a positive SKR, owing to the reduced thermal noise, which enables secure key generation even at elevated temperatures. It should be noted that Figure~\ref{fig4} presents results only for the PMCS protocol. As shown in equations (\ref{29}) and (\ref{30}), PMCS typically introduces more noise than GMCS, leading to a more conservative performance estimate. Consequently, within the parameter ranges where PMCS achieves a positive SKR, the GMCS protocol is expected to achieve no worse performance under the same parameter settings.

\begin{figure*}[ht]
	\subfloat[ $f=10 \mathrm{THz}$]{\includegraphics[width=0.3\textwidth]{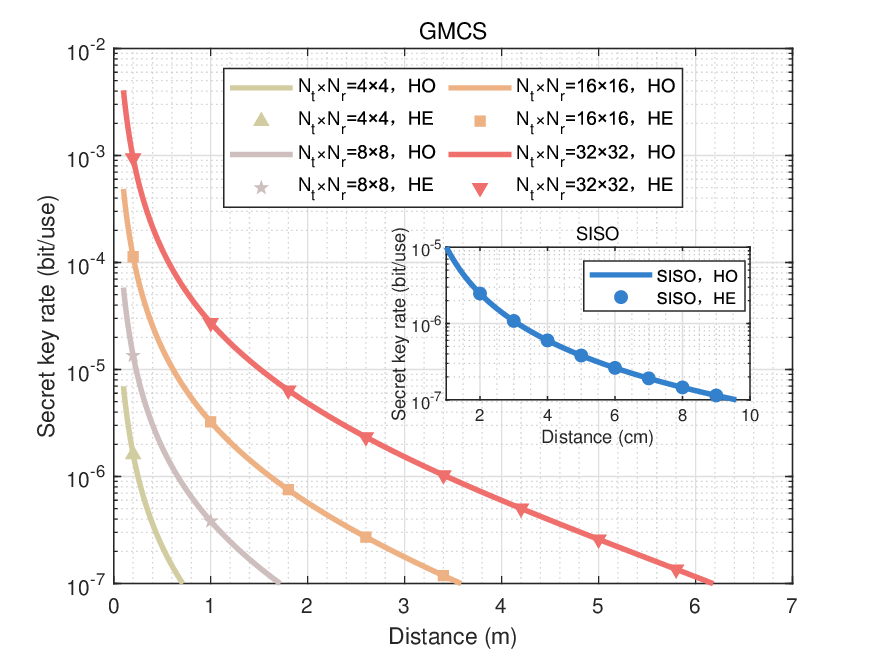}}
	\hfill 	
	\subfloat[$f=15 \mathrm{THz}$]{\includegraphics[width=0.3\textwidth]{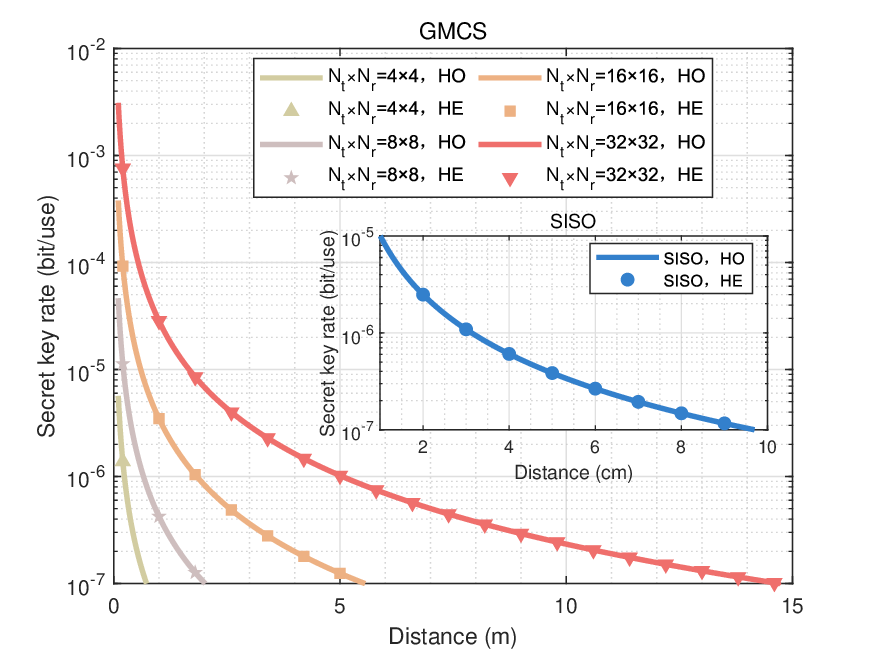}}
	\hfill 	 
	\subfloat[$f=30 \mathrm{THz}$]{\includegraphics[width=0.3\textwidth]{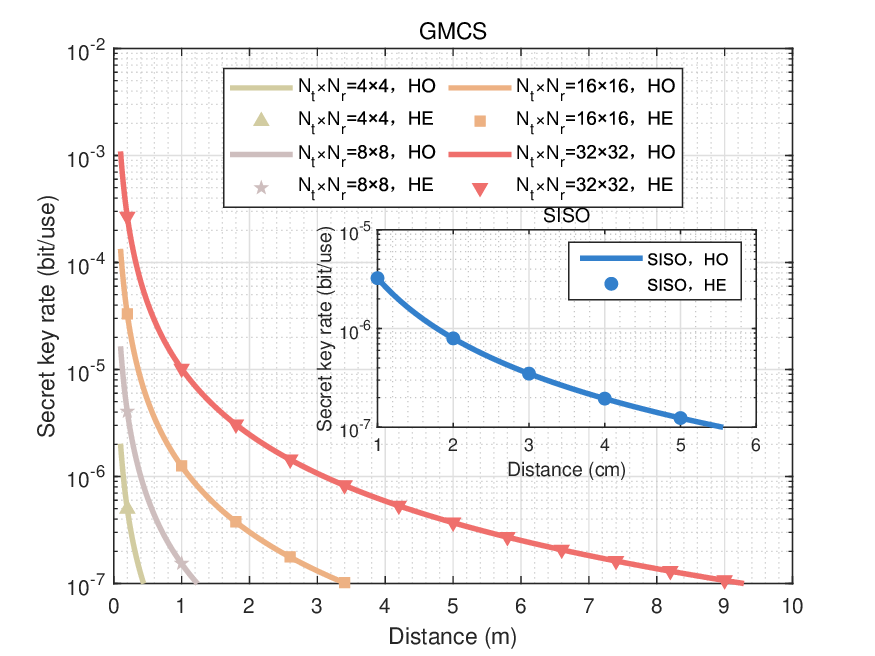}}
	\hfill	
	\subfloat[ $f=10 \mathrm{THz}$]{\includegraphics[width=0.3\textwidth]{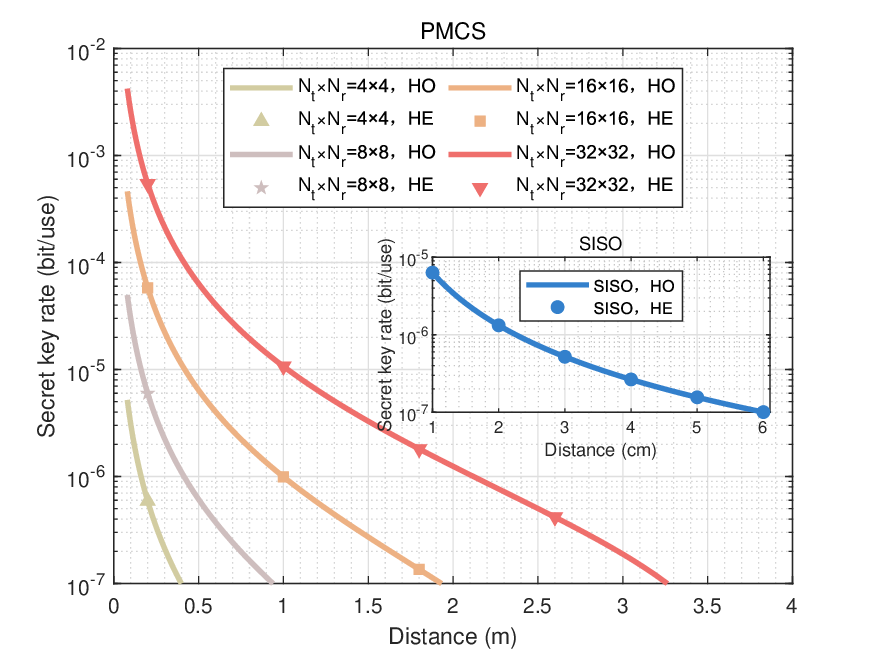}}
	\hfill 	
	\subfloat[$f=15 \mathrm{THz}$]{\includegraphics[width=0.3\textwidth]{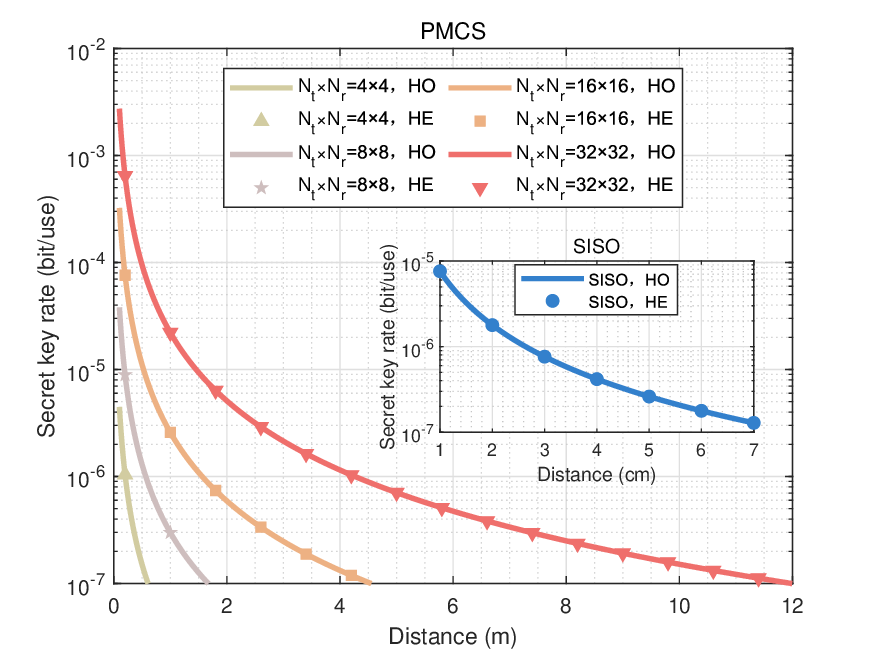}}
	\hfill 	 
	\subfloat[$f=30 \mathrm{THz}$]{\includegraphics[width=0.3\textwidth]{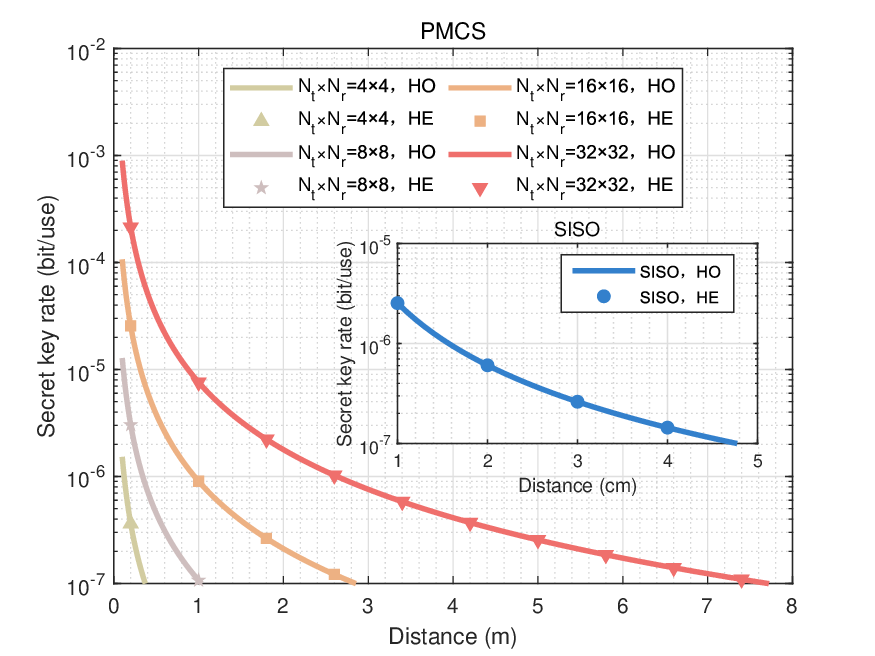}}
	\hfill	
	\caption{Two-dimensional diagram of SKR versus transmission distance for low-configuration MIMO at different carrier frequencies under the GMCS (a),(b),(c) and PMCS (d),(e),(f) protocols. HO: homodyne, HE: heterodyne detection. $N_a=0.005$,  temperature {$T_e$}$=300$ K, $n=2$.}
	\label{fig5}
\end{figure*}

Figure \ref{fig5} analyzes the impact of low-order MIMO configurations on the performance of the GMCS and PMCS protocols across different channel frequencies at room temperature, with baseline SISO systems included for comparison. When comparing curves of the same color, the  proposed protocol achieves the longest transmission distance with a carrier frequency  $f=15$ THz under identical MIMO configurations, primarily because of its lower atmospheric absorption and reduced thermal noise. Consequently,  in the following studies, we fix the carrier frequency at $f = 15$ THz. When comparing curves of different colors, the benefit of MIMO becomes evident in improving both the SKR and the transmission distance. Specifically, the GMCS protocol at 15 THz increases the maximum transmission distance from approximately 0.76 m to 2.11 m, 5.80 m, and 14.99 m when the antenna array is increased from $N_t \times N_r = 4 \times 4$ to $8 \times 8$, $16 \times 16$, and $32 \times 32$. For the PMCS protocol under the same  frequency and configurations, the corresponding maximum distances are 0.60 m, 1.67 m, 4.54 m, and 11.96 m. Although its performance is inferior to the GMCS protocol, the $32 \times 32$ MIMO configuration remains sufficient for short- to medium-range indoor wireless quantum networks. It is interesting to note that homodyne and heterodyne detection exhibit remarkably similar SKR performance.  Although heterodyne detection can measure two orthogonal quadratures simultaneously, it also modifies the corresponding Holevo information available to Eve. Under the considered parameter settings, these effects partially offset each other, resulting in comparable macroscopic SKR performance between homodyne and heterodyne detection.

\begin{figure*}[t]
	\subfloat[$n=2$]{\includegraphics[width=0.24\textwidth]{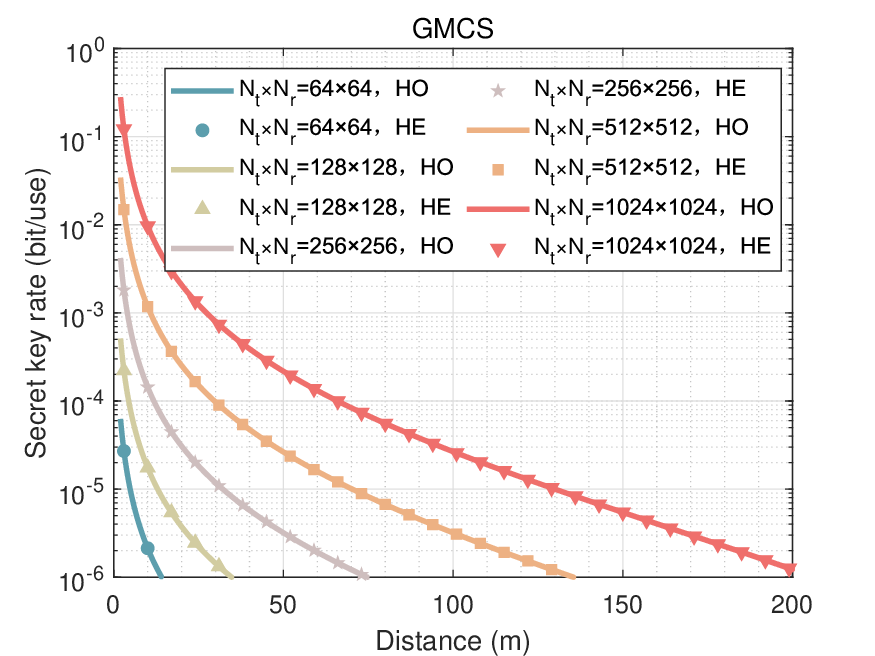}}
	\hfill 	
	\subfloat[$n=3$]{\includegraphics[width=0.24\textwidth]{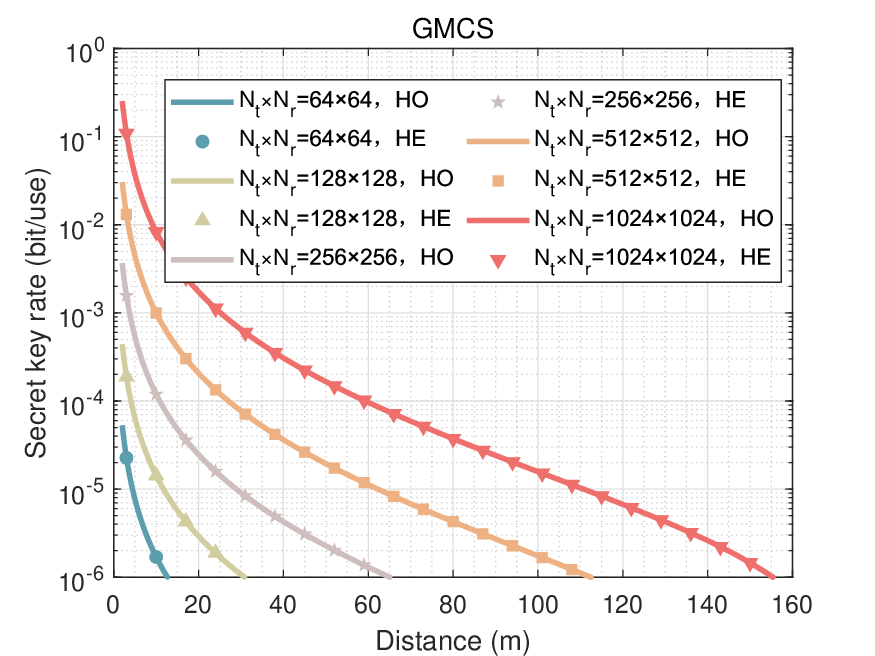}}
	\hfill 	
	\subfloat[$n=4$]{\includegraphics[width=0.24\textwidth]{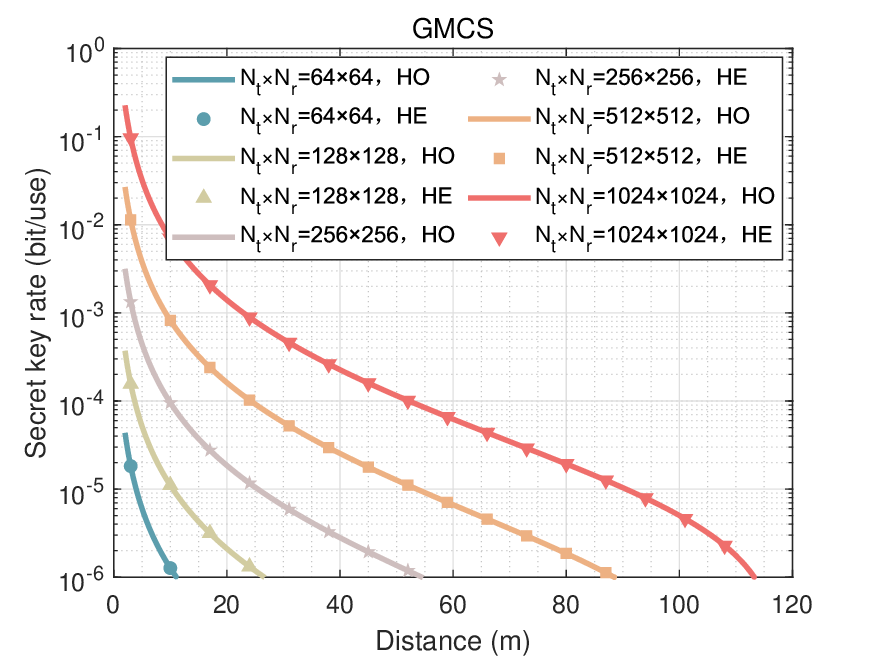}}
	\hfill	
	\subfloat[$n=5$]{\includegraphics[width=0.24\textwidth]{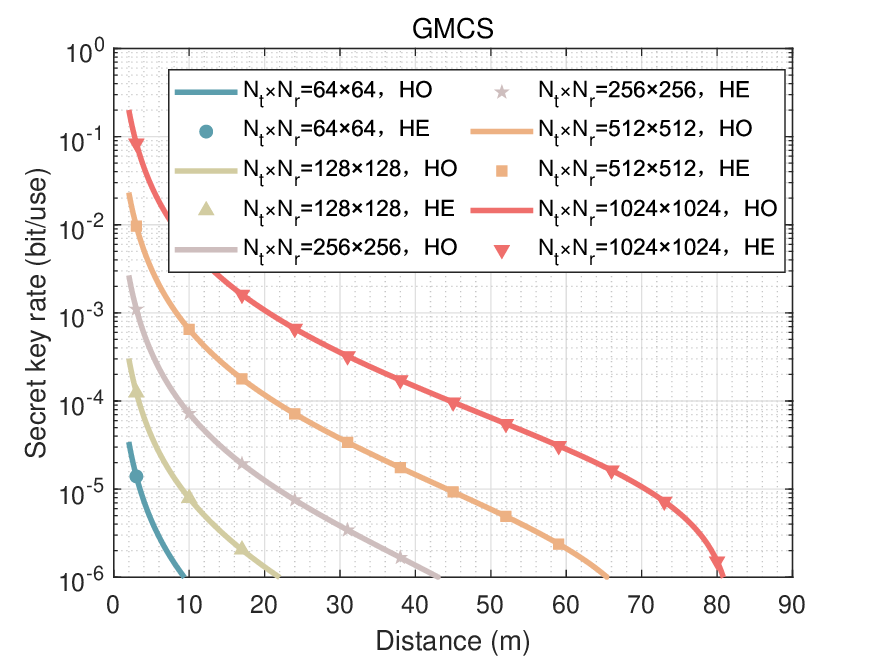}}
	\hfill 	
	\newline
	\subfloat[$n=2$]{\includegraphics[width=0.24\textwidth]{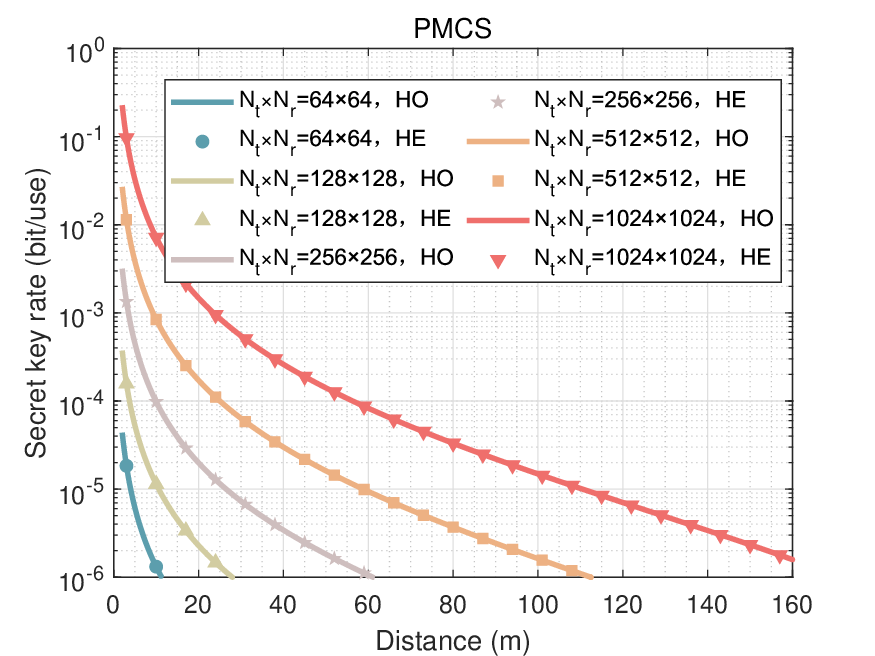}}
	\hfill 	
	\subfloat[$n=3$]{\includegraphics[width=0.24\textwidth]{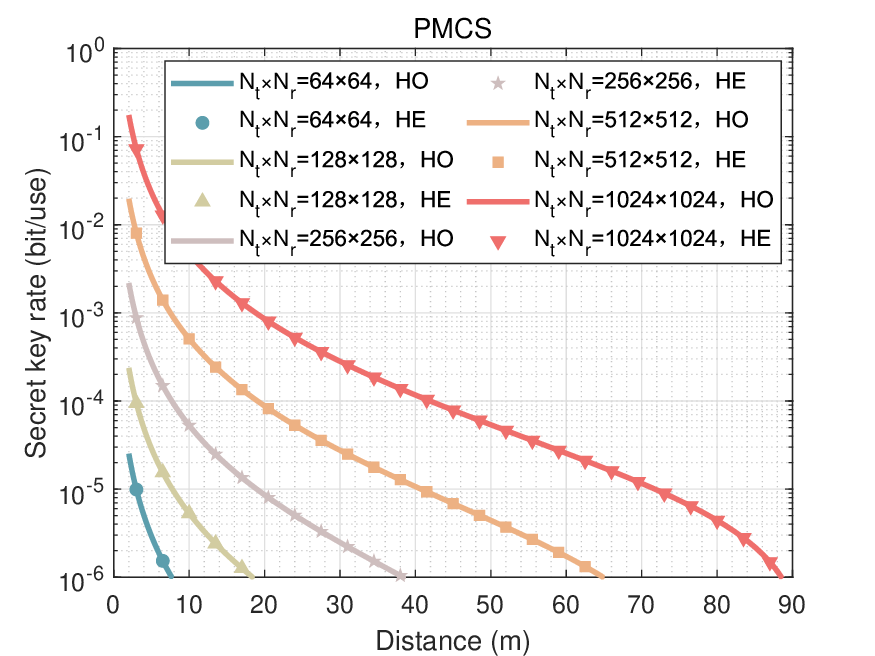}}
	\hfill 	
	\subfloat[$n=4$]{\includegraphics[width=0.24\textwidth]{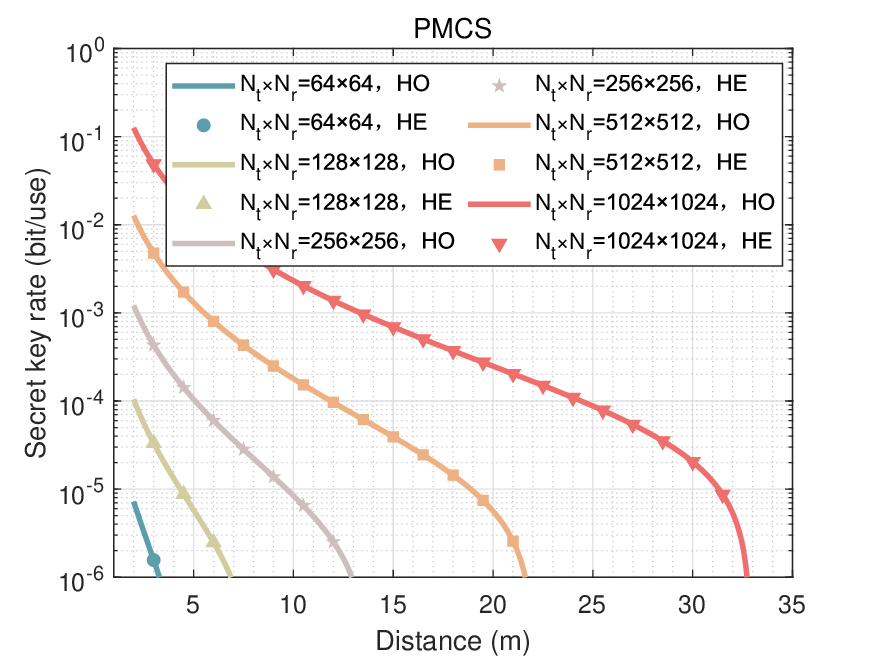}}
	\hfill	
	\subfloat[$n=5$]{\includegraphics[width=0.24\textwidth]{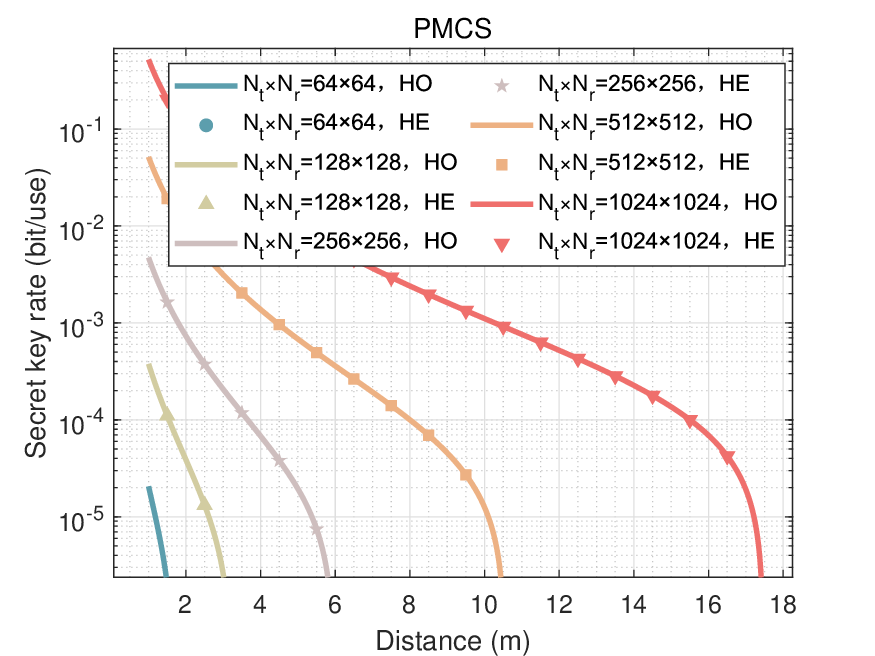}}
	\hfill 	
	\caption{SKR as a function of transmission distance using GMCS (a),(b),(c),(d) and PMCS (e),(f),(g),(h) protocols under high-configuration MIMO setups with varying numbers of users. HO: homodyne, HE: heterodyne detection. $N_a=0.005$, $f=15$ THz, {$T_e$}$=300$ K.}
	\label{fig6}
\end{figure*}

\begin{figure*}[t]
	\subfloat[$N_t\times N_r=32\times32$]{\includegraphics[width=0.48\textwidth]{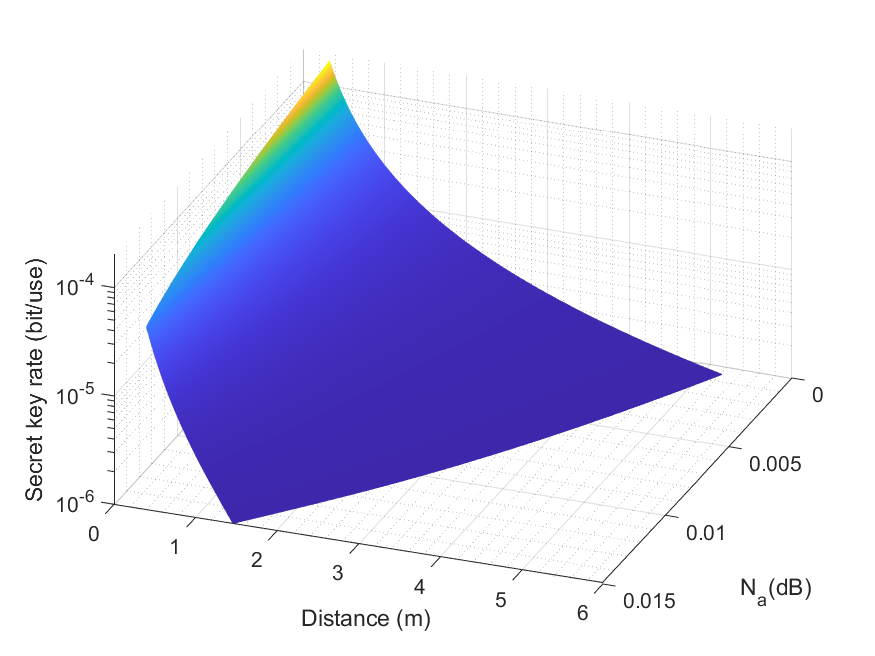}}
	\hfill 	
	\subfloat[$N_t\times N_r=1024\times1024$]{\includegraphics[width=0.48\textwidth]{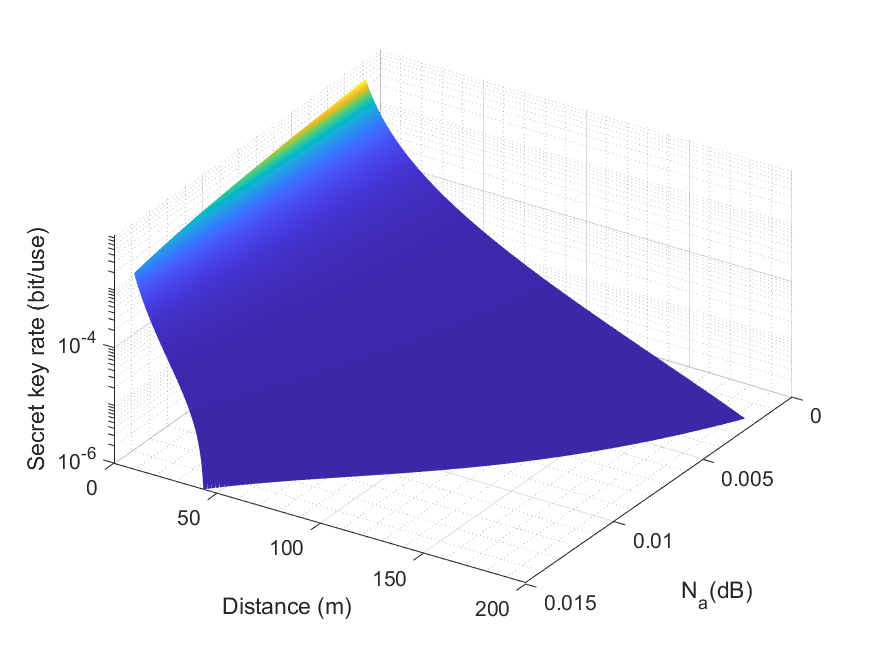}}
	\hfill 	
	\caption{The 3D diagram illustrates the relationship between homodyne SKR, transmission distance, and  attenuator noise for the PMCS protocol under both low- (a) and high-configuration (b) MIMO. $f=15$ THz,  {$T_e$}$=300$ K, $n=2$. }
	\label{fig7}
\end{figure*}
Figure~\ref{fig6} illustrates the performance of the GMCS and PMCS protocols at a carrier frequency of 15~THz in terms of SKR and transmission distance, for user numbers $n = 2, 3, 4, 5$, under high-order antenna configurations ranging from $N_t \times N_r = 64 \times 64$ to $1024 \times 1024$. The results show that large-scale MIMO significantly enhances the maximum transmission distance. For instance, when users number $n=2$ and $N_t \times N_r =1024 \times 1024$ antenna array, GMCS and PMCS achieve maximum distances of approximately 200~m and 160~m, respectively, whereas $N_t \times N_r =64 \times 64$ antenna configuration only reaches 18~m and 16~m. This improvement is primarily due to the ability of large-scale MIMO to establish multiple parallel SISO channels, enabling simultaneous generation of multiple secret-key streams over parallel SISO subchannels. {Under the adopted idealized assumptions, these results indicate the potential of large-scale MIMO arrays to improve the transmission distance of THz CV-QSS systems in short-range outdoor scenarios.}
\begin{figure*}[t]
	\subfloat[ $N_{t} \times N_{r} =32\times32$]{\includegraphics[width=0.48\textwidth]{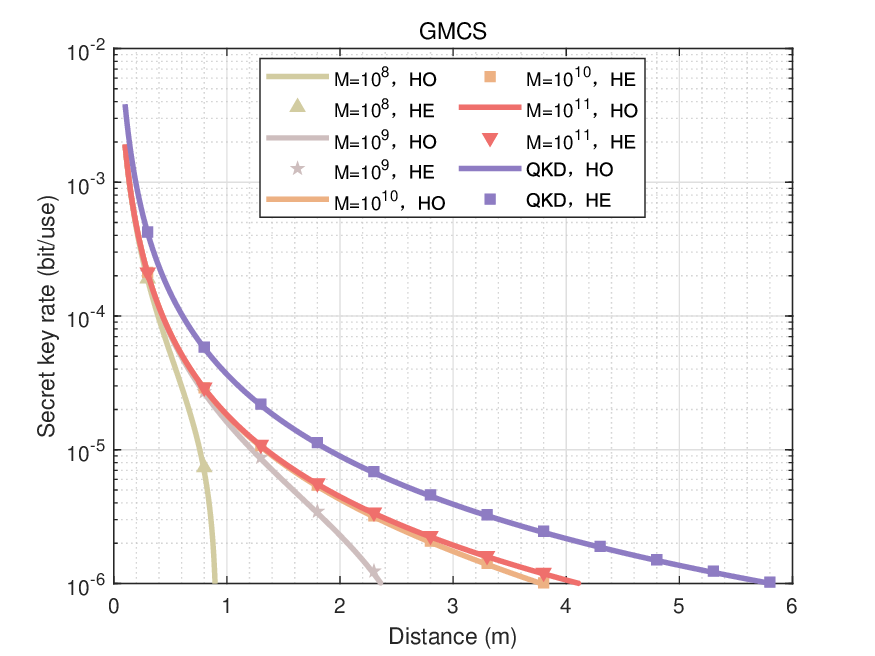}}
	\hfill 	
	\subfloat[$N_{t} \times N_{r}=1024\times1024$]{\includegraphics[width=0.48\textwidth]{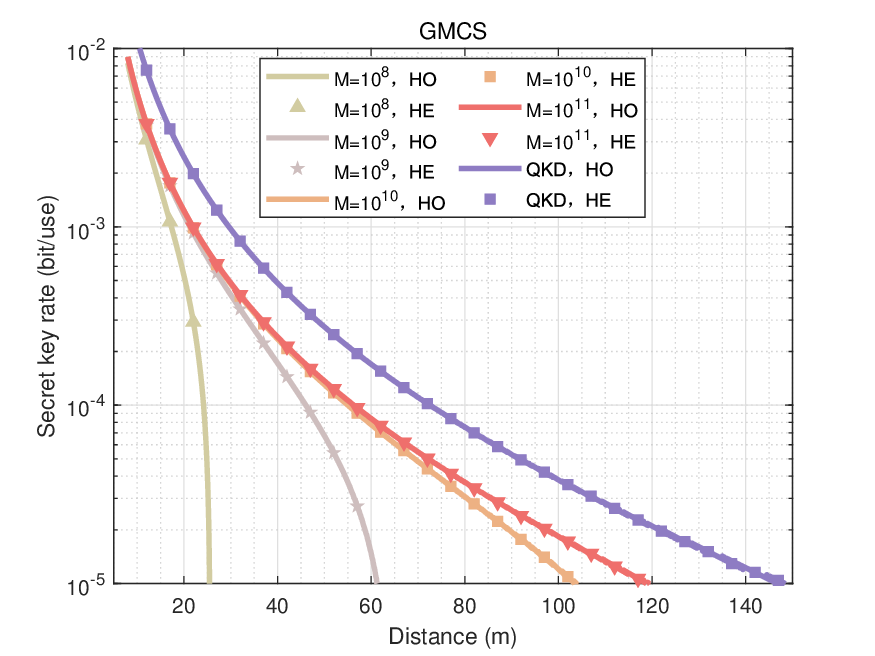}}
	\hfill 	
	\subfloat[$N_{t} \times N_{r}=32\times32$]{\includegraphics[width=0.48\textwidth]{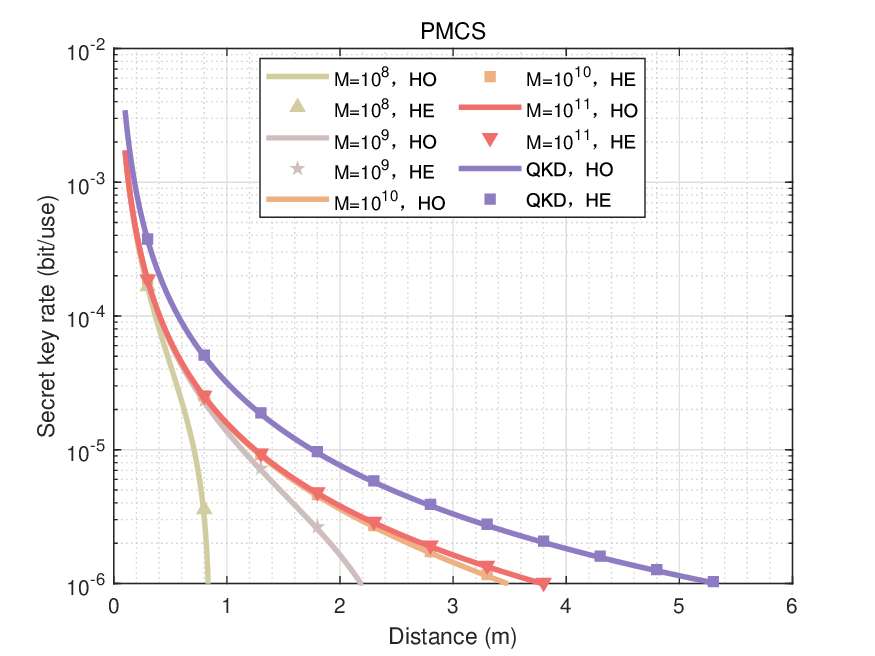}}
	\hfill	
	\subfloat[$N_{t} \times N_{r}=1024\times1024$]{\includegraphics[width=0.48\textwidth]{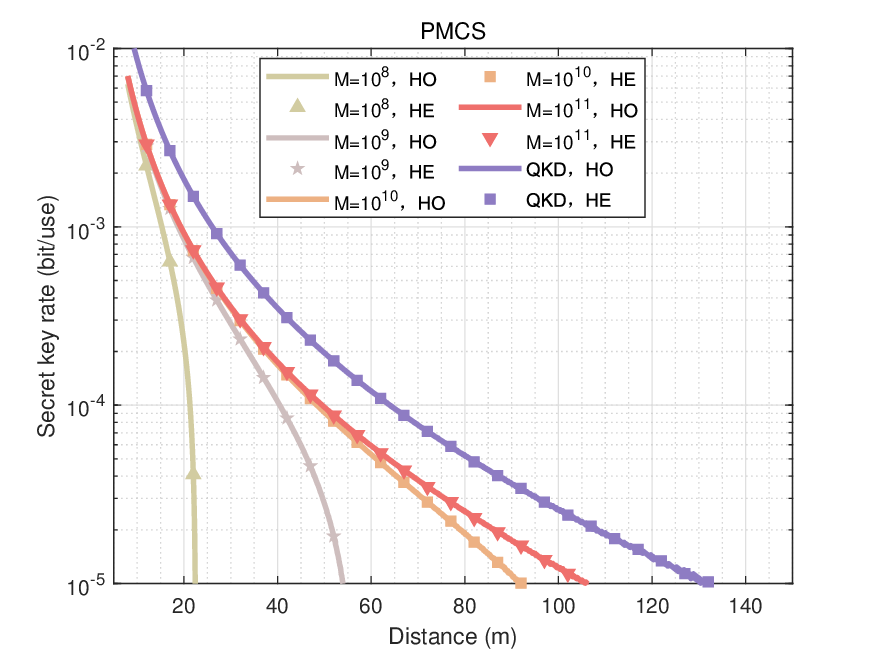}}
	\hfill 	
	\caption{Relationship between { composable} finite-size SKRs and transmission distances for GMCS (a),(b) and PMCS (c),(d) protocols in high-configuration and low-configuration MIMO systems at different block sizes. HO: homodyne, HE: heterodyne detection. $N_a=0.005$, $f=15$ THz, {$T_e$}$=300$ K, $O=N = M/2$, $n=2$.}
	\label{fig9}
\end{figure*}

However, the same-color curves show that the maximum transmission distance decreases as the number of users increases. Specifically, when the number of users rises from 2 to 5, the maximum distances for GMCS and PMCS with the $N_t \times N_r =1024 \times 1024$ array drop to approximately 80.8~m and 17~m, respectively. For the GMCS protocol, this reduction is mainly caused by additional untrusted noise introduced by new users through their local devices or controlled channels (see equations~(\ref{188}) and (\ref{29})). For the PMCS protocol, the decrease is attributed not only to the additional noise associated with newly introduced users, but also to the extra noise introduced by the attenuator during passive state preparation (see equation~(\ref{30})). These findings highlight a fundamental trade-off in multi-party QSS systems: although larger MIMO arrays improve both transmission distance and SKR, overall system performance declines as the network scales. Therefore, {the number of users and the antenna configuration should be jointly optimized to balance system scalability and performance}.

The preceding results indicate that the performance gap between GMCS and PMCS protocols is primarily due to the excess noise introduced by the  attenuator in the PMCS scheme.  To further investigate this effect, Figure \ref{fig7} is used to present the impact of  attenuator noise on both the {homodyne} SKR and the maximum transmission distance of the PMCS protocol. {Similar conclusions can also be obtained for the heterodyne detection case.} The left subfigure corresponds to the low-configuration antenna setup with $N_t \times N_r = 32 \times 32$, while the right subfigure represents the high-configuration antenna setup with $N_t \times N_r = 1024 \times 1024$. The results indicate that as the  attenuator noise increases, the maximum achievable secure transmission distance decreases noticeably for all antenna configurations. These findings highlight the sensitivity of the PMCS protocol to excess noise and emphasize the importance of controlling attenuator noise in implementation-oriented system designs. Therefore, designing PMCS-based CV-QSS systems requires careful management of  attenuator noise to avoid significant performance degradation.

In the asymptotic scenario, the SKR is evaluated by assuming an asymptotically large block length. This provides an upper bound on both the achievable rate and the transmission distance. However, such an ideal case is not practical. In real systems, { composable} finite-size effects must be considered. Under { composable} finite-size conditions, the raw block length is limited, and {a portion of the raw data must be sacrificed for parameter estimation, and finite-size corrections associated with privacy amplification must be included.} Meanwhile, it is assumed that during parameter estimation, user 1 and the dealer always take the lower bound of the channel transmittance and the upper bound of the total excess noise. In our simulations, the number of samples used for parameter estimation is set to half of the total data size, i.e.,  $O = N = M/2$.

Figure~\ref{fig9} shows the { composable} finite-size SKR as a function of transmission distance for the GMCS and PMCS protocols under both low- and high-order MIMO architectures at a carrier frequency of 15~THz. For comparison, the corresponding one-user  MIMO THz CV-QKD is included as a theoretical upper bound. In all cases, larger block lengths yield higher SKRs and longer transmission distances because they reduce parameter-estimation uncertainty. As a result, the conservative lower bound on the channel transmittance and the conservative upper bound on the excess noise become less pessimistic, leading to improved finite-size SKR performance (see equations~(\ref{52}) and (\ref{53})). Moreover, {the finite-size penalties associated with privacy amplification and the asymptotic equipartition property (AEP) are correspondingly reduced.} For instance, with a $N_t \times N_r = 32 \times 32$ antenna array and a block length of $10^{11}$, the GMCS protocol achieves a secure transmission distance of approximately 5.9~m, which is sufficient for small-scale indoor wireless communication scenarios.  {Under a larger MIMO configuration of $N_t \times N_r = 1024 \times 1024$ and a block length of $10^{10}$, the PMCS protocol achieves a transmission distance of approximately 105~m under idealized assumptions. This result suggests its potential performance advantage in short-range outdoor THz quantum communication scenarios. Moreover, the observed trade-off between antenna scale and block length indicates that these two parameters should be jointly optimized to achieve the desired system SKR while limiting the associated resource overhead.}

\begin{figure}[t]
	\centering
	\includegraphics[width=\linewidth]{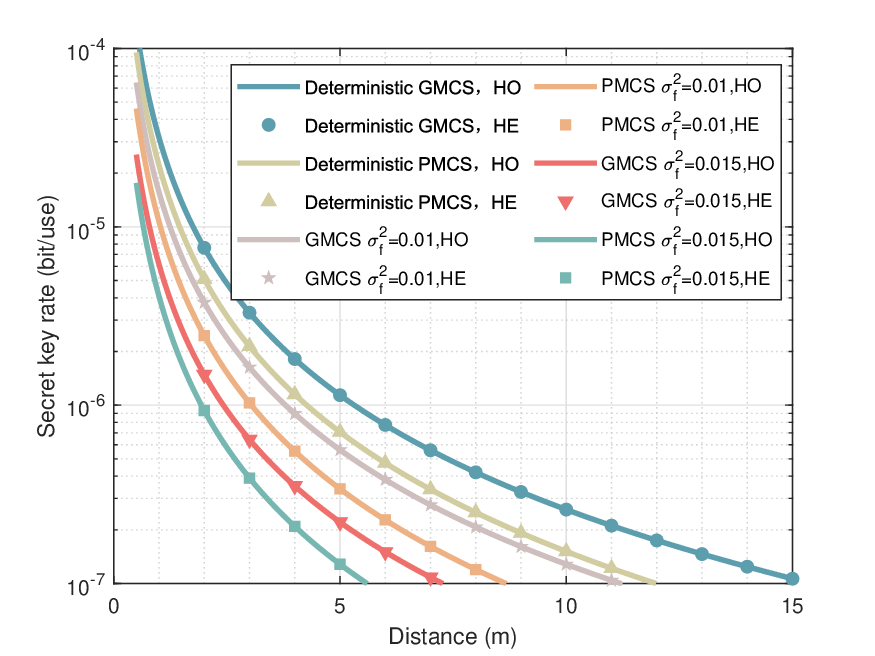}
	\caption{Asymptotic SKR performance of the proposed MIMO-assisted THz QSS protocol under different fluctuation intensities. HO: homodyne, HE: heterodyne detection. $N_a=0.005$, $f=15$ THz, {$T_e$}$=300$ K, $n=2$.}
	\label{fig100}
\end{figure}

{In the previous simulations, the transmittance of each decomposed SISO eigenchannel is obtained from the deterministic MIMO channel model. To evaluate the impact of stochastic channel fluctuations on the proposed MIMO-assisted THz QSS protocol, we introduce a Monte Carlo-based random fading simulation. Specifically, the NLoS eigenchannel transmittance is modeled as $\hat{T}_{k_s}\sim\mathcal{N}(T_{k_s},\sigma_f^2)$, where $s=2,3,...,r_{\mathrm{min}}$, $T_{k_s}$ denotes the deterministic transmittance obtained from the original channel model, and $\sigma_f^2$ characterizes the fluctuation intensity of the fading channel. For each transmission distance, $10^4$ independent Monte Carlo channel realizations are generated to estimate the statistical expectation of the random transmittance, denoted by $\mathbb{E}(\hat{T}_{k_s})$. Meanwhile, the fading-induced excess noise is modeled as $V_0[\mathbb{E}(\hat{T}_{k_s})-(\mathbb{E}(\sqrt{\hat{T}_{k_s}}))^2]$\cite{25}. Figure \ref{fig100} shows that the maximum transmission distance and asymptotic SKR decrease as the fluctuation intensity increases.}

 \section{CONCLUSION}
  \subsection{Contributions}
This paper presents a MIMO THz $(n,n)$ threshold CV-QSS protocol {designed to support secure multi-user key sharing over THz free-space channels.} In the proposed scheme, decryption requires the collaboration of all users’ secret-key shares, which prevents any subset of fewer than $n$ users from reconstructing the secret. The communication procedure is described in detail, and the SKRs of eight protocol variants are derived. At the transmitter, Gaussian and passive modulation are employed to generate coherent states, while at the receiver, homodyne and heterodyne detection are adopted for quantum state measurement. Both asymptotic and { composable finite-size analyses are carried out to obtain the theoretical upper bounds and the corresponding finite-resource performance under the stated assumptions.} These analyses provide a more comprehensive theoretical basis for evaluating the security performance of the proposed scheme under the stated assumptions.

Simulation results demonstrate that MIMO technology can effectively mitigate free-space path loss in the considered QSS protocol and enables key distribution among multiple users. Specifically, low-order antenna configurations show potential for short-range indoor networks with a limited number of users, whereas high-order MIMO configurations provide theoretical beamforming and spatial gains under idealized assumptions, suggesting potential performance improvements for short-range outdoor multi-user THz scenarios.  Both the GMCS QSS protocol based on AM and PM, and the PMCS QSS protocol employing an attenuator and a detector, exhibit distinct advantages and limitations. {The GMCS QSS protocol achieves higher theoretical SKRs owing to its lower excess noise, indicating its potential advantage in scenarios where high SKRs and longer transmission distances are desired under idealized conditions. In contrast, the PMCS protocol, with its simpler state-preparation architecture, provides a useful alternative for short-range CV-QSS designs.} Furthermore,  { composable finite-size SKR simulation reveals that increasing the block size significantly improves parameter estimation accuracy, enabling higher SKRs and longer secure transmission distances. This provides a theoretical reference for understanding the impact of finite-size effects on the proposed MIMO-assisted THz CV-QSS scheme.} 

 \subsection{Limitations}

Due to the theoretical focus of this study, the current analysis assumes perfect {channel state information (CSI)}, {ideal beamforming}, { perfect mode matching}, and {stable phase synchronization}. Practical impairments such as  imperfect {phase-reference sharing, synchronization errors}, beam misalignment, blockage, user mobility, { mode mismatch}, and hardware impairments are not explicitly considered. {Moreover, this paper primarily focuses on deterministic LoS and NLoS channel models, where beamforming and SVD are employed to decompose the MIMO channel into parallel SISO subchannels. Although a simplified Monte Carlo-based Gaussian fading simulation is introduced to evaluate the sensitivity of the proposed scheme to random transmittance fluctuations, a rigorous statistical fading-channel model would require a more comprehensive characterization of practical THz MIMO propagation, including the statistical distributions of fading, spatial correlations among eigenchannels, imperfect CSI, mode mismatch, and hardware non-idealities.} {Future work will introduce stochastic THz MIMO channel models, with particular emphasis on statistical fading distributions, imperfect CSI, mode mismatch, and hardware non-idealities, to further evaluate the robustness of the proposed MIMO THz CV-QSS scheme under non-ideal operating conditions.}


\appendices

\section{The atmospheric channel model}\label{appendixa}
 The atmospheric channel matrix between user $k$, and the next user is defined as \cite{51}: 
\begin{equation}\mathbf{H_{k}}=\sum_{l=1}^{L_{k}}\sqrt{\gamma_{l}}e^{j2\pi f\tau_{l}}\psi_{N_{r_{k+1}}}(\phi_{l}^{r,k+1})\psi_{N_{t_k}}^{\dagger}(\phi_{l}^{t,k}),\end{equation}
where $L_k$ is the number of propagation paths between $k$-th user and ($k+1$)-th user. $\phi_{l}^{t,k}$ and $\phi_{l}^{r,k+1}$ are the angle of departure of user $k$ and angle of arrival of user $k+1$ in $l$-th path on the uniform linear array. $\tau_{l}$ and $\gamma_{l}$  denote propagation delay and path loss of the $l$-th multipath in atmospheric channel. $\psi_{N_{r_{k+1}}}$ and $\psi_{N_{t_k}}$ denote the array response vectors of the uniform linear arrays at the receiver of user $k+1$ and the transmitter of user $k$, respectively, which are defined as \cite{52}:
\begin{equation}\small
	\begin{aligned}
		&\psi_{N_{r_{k+1}}}\left(\theta\right)=\frac{1}{\sqrt{N_{r_{k+1}}}}\left[1,e^{j\frac{2\pi}{\lambda}d_{a}\sin\theta},\ldots,e^{j\frac{2\pi}{\lambda}d_{a}(N_{r_{k+1}}-1)\sin\theta}\right]^{T},\\&\psi_{N_{t_k}}\left(\theta\right)=\frac{1}{\sqrt{N_{t_k}}}\left[1,e^{j\frac{2\pi}{\lambda}d_{a}\sin\theta},\ldots,e^{j\frac{2\pi}{\lambda}d_{a}(N_{t_k}-1)\sin\theta}\right]^{T},
	\end{aligned}
\end{equation}
where $d_a$ represents {the antenna spacing distance}.  We denote the rank of the channel matrix $\mathbf{H_k}$ as $r_k$, which represents the number of parallel SISO subchannels into which the MIMO channel between user $k$ and the subsequent user can be decomposed. Within the CV-QSS protocol framework, the maximum number of parallel SISO subchannels that can be simultaneously established is determined by the minimum channel rank, given by
\begin{equation}
r_{\mathrm{min}} = \min\{r_1, r_2, \ldots, r_n\}.
\end{equation}

 \section{Proof of Proposition 2}\label{appendixb}
To provide a conservative and experimentally meaningful assessment of the maximum transmission distance and SKR, we evaluate the  { composable finite-size SKR of the MIMO CV-QSS protocol} {under Gaussian collective attacks}  \cite{53}. Then, the  composable finite-size SKR, denoted by $R_{\mathrm{MIMO}}^\mathrm{F}$, is given by
{
\begin{equation}\small
	R_{\mathrm{MIMO}}^\mathrm{F}=\sum_{j=1}^{r_{\mathrm{min}}}\left(\frac{N}{M} R_{\mathrm{SISO}_{1_j}}^\mathrm{F}-\frac{\sqrt{N}}{M}\Delta_{AEP}-\Delta_{PA}\right),
\end{equation}}
where $R_{\mathrm{SISO}_{1_j}}^\mathrm{F} = \beta I (U_{1_j}:D_{1_j}) -  \chi^{\epsilon_{\mathrm{PE}}}(D_{1_j}:E)$. {The parameters $\beta$} and $I(U_{1_j}:D_{1_j})$ follow the same definitions as in equation (\ref{6}). The term $\chi^{\epsilon_{\mathrm{PE}}}(D_{1_j}:E)$ denotes the maximum Holevo information accessible to the eavesdropper with failure probability $\epsilon_{\mathrm{PE}}$ \cite{53,54}. Here, $N$ is the number of data used to generate the shared key between user 1 and the dealer, while $M$ represents the total number of transmitted data. The remaining data, denoted as $O = M - N$, is reserved for parameter estimation. The $\Delta_{PA}$ associated with the security of privacy amplification is defined as { \cite{53,55}:
 \begin{equation}\Delta_{PA}=2\frac{\log_2(1/(2\bar{\epsilon}))}{M},\end{equation}}
 where $\bar{\epsilon}$ is thet  privacy amplification parameter \cite{43}. $\Delta_{AEP}$ denotes the finite-size correction arising from the AEP, which is given by  { \cite{53}
 \begin{equation}\begin{aligned}
 		& \Delta_{AEP}=36+24\sqrt{\log_{2}(2/\epsilon_{\mathrm{sm}}^{2})}+ \\
 		& 2\log_{2}(2/(\epsilon^{2}\epsilon_{\mathrm{sm}}))+20\epsilon_{\mathrm{sm}}/(\epsilon\sqrt{N}),
 \end{aligned}\end{equation}}
where {the overall security parameter $\epsilon{=}2\epsilon_{\mathrm{sm}}+\bar{\epsilon}{+}\epsilon_{\mathrm{PE}}{+}\epsilon_{\mathrm{cor}}$ and  the smoothing parameter $\epsilon_{\mathrm{sm}}$ is associated with the smooth min-entropy estimation, $\epsilon_{\mathrm{cor}}$ represents the maximum failure probability associated with the error correction process.} { In this work, the parameters are set to $\epsilon_{\mathrm{sm}}=\bar{\epsilon}=\epsilon_{\mathrm{PE}}=\epsilon_{\mathrm{cor}}=10^{-10}$.}  To evaluate the { composable} finite-size SKR, the channel transmittance $T_{1_j}$ and the total excess noise $\hat{\xi}_{1_j}$ must first be estimated. Their estimation is based on $O$ correlated pairs $(x_i, y_i)$, with $i = 1,2, \ldots, O$. Under the normal distribution model, the relationship between user1’s and dealer’s data can be expressed as:
 \begin{equation}y_{i}=t_{1_j}x_{i}+z_{1_j},\end{equation}
where $t_{1_j}=\sqrt{T_{1_j}}$ and $z_{1_j}\sim\mathcal{N}(0,\sigma_{1_j}^{2})$. Here $\sigma_{1_j}^{2}=1+T_{1_j}\hat{\xi}_{1_j}$ and $\hat{\xi}_{1_j}$ represents the total channel excess noise in the $j$-th SISO channel between the first user and the dealer. 
Under this linear regression model, the maximum likelihood estimators $\hat{t}_{1_j}$ and $\hat{\sigma}_{1_j}^2$ are given by
\begin{equation}\label{33}
	\hat{t}_{1_j}=\frac{\sum_{k=1}^{O}x_{k}y_{k}}{\sum_{k=1}^{O}x_{k}^{2}},\quad\hat{\sigma}_{1_j}^2=\frac{1}{O}\sum_{k=1}^{O}(y_{k}-\hat{t}_{1_j}x_{k})^{2}.\end{equation}

 Additionally, $\hat{t}_{1_j}$ and $\hat{\sigma}_{1_j}^2$ are independent estimators, with distributions given by:
\begin{equation}\label{34}
	\hat{t}_{1_j}\sim\mathcal{N}\left(t_{1_j},\frac{\sigma_{1_j}^2}{\sum_{i=1}^Ox_i^2}\right)\quad\mathrm{and}\quad\frac{O\hat{\sigma}_{1_j}^2}{\sigma_{1_j}^2}\sim\chi^2(O-1),\end{equation}

{Based on this information, the user 1 and dealer  evaluate these parameters with following confidence intervals:
\begin{equation}\begin{aligned}
	&
	t_{1_j,e}\in[\hat{t}_{1_j}-\Delta(t_{1_j}),\hat{t}_{1_j}+\Delta(t_{1_j})],\\
	&  \sigma_{1_j,e}^{2}\in[\hat{\sigma}_{1_j}^{2}-\Delta(\sigma_{1_j}^{2}),\hat{\sigma}_{1_j}^{2}+\Delta(\sigma_{1_j}^{2})].\end{aligned}\end{equation}
where
\begin{equation}\begin{aligned}
		& \Delta(t_{1_j})=z_{\epsilon_{\mathrm{PE}/2}}\sqrt{\frac{\hat{\sigma}_{1_j}^{2}}{OV_0}}, \\
		& \Delta(\sigma_{1_j}^{2})=z_{\epsilon_{\mathrm{PE}}/2}\frac{\hat{\sigma}_{1_j}^{2}\sqrt{2}}{\sqrt{O}}.
\end{aligned}\end{equation}}
where $z_{\epsilon_{\mathrm{PE}}/2}=6.5$ and $\hat{t}_{1_j}$ and $\hat{\sigma}_{1_j}^2$ are set to their respective expected values: $E[\hat{t}_{1_j}]=\sqrt{T_{1_j}}$ and $E[\hat{\sigma}_{1_j}^2]=1+T_{1_j}\hat{\xi}_{1_j}$. For the adopted parameter ranges, numerical verification shows that the following monotonicity relations hold:

 \begin{equation}\label{31}
	\left.\frac{\partial \chi^{\epsilon_{\mathrm{PE}}}(D_{1_j}:E)}{\partial t_{1_j}}\right|_{\sigma_{1_j}^2}<0\quad\mathrm{and}\quad\left.\frac{\partial \chi^{\epsilon_{\mathrm{PE}}}(D_{1_j}:E)}{\partial\sigma_{1_j}^2}\right|_{t_{1_j}}>0.\end{equation}
To obtain the most conservative SKR, the Holevo bound is maximized by adopting the worst-case estimation strategy, i.e., by choosing the minimum  $t_{1_j,\mathrm{min}}$ and the maximum  $\sigma_{1_j, \mathrm{max}}^2$ within their corresponding confidence intervals. After simplification, we obtain:
\begin{equation}\label{36}\begin{aligned}
		& 	t_{1_j,\mathrm{min}}\approx\sqrt{T_{1_j}}-6.5\sqrt{\frac{1+T_{1_j}\hat{\xi}_{1_j}}{OV_0}}, \\
		& \sigma_{1_j, \mathrm{max}}^2\approx1+T_{1_j}\hat{\xi}_{1_j}+6.5\frac{\sqrt{2}(1+T_{1_j}\hat{\xi}_{1_j})}{\sqrt{O}}.
\end{aligned}\end{equation}
By substituting $\sigma_{1_j, \mathrm{max}}^2=1+T_{1_j}\hat{\xi}_{1_j, \mathrm{max}}$,  $t_{1_j,\mathrm{min}}=\sqrt{T_{1_j,\mathrm{min}}}$, equations (\ref{29}) and (\ref{30}), for GMCS protocol, we have

\begin{equation}\label{52}\begin{aligned}
		& 	T_{1_j,\mathrm{min}}^G=(\sqrt{T_{1_j}}-6.5\sqrt{\frac{1+T_{1_j}\sum_{k=1}^n\xi_{k_j}}{OV_0}})^{2}, \\
			& \hat{\xi}_{1_j, \mathrm{max}}^G=\hat{\xi}_{1_j}+6.5\frac{(1+T_{1_j}\sum_{k=1}^n\xi_{k_j})\sqrt{2}}{T_{1_j}\sqrt{O}}.
	\end{aligned}\end{equation}

For PMCS protocol, we have
\begin{equation}\label{53}\begin{aligned}
		& 	T_{1_j,\mathrm{min}}^P=(\sqrt{T_{1_j}}-6.5\sqrt{\frac{1+T_{1_j}\sum_{k=1}^n(\xi_{k_j}+N_{k_j})}{OV_0}})^{2}, \\
		& \hat{\xi}_{1_j, \mathrm{max}}^P=\hat{\xi}_{1_j}+6.5\frac{(1+T_{1_j}\sum_{k=1}^n(\xi_{k_j}+N_{k_j}))\sqrt{2}}{T_{1_j}\sqrt{O}}.
\end{aligned}\end{equation}
By replacing $T_{1_j}$ and $\hat{\xi}_{1_j}^G$ in equation (\ref{12}) with $T_{1_j,\mathrm{min}}^G$ and $\hat{\xi}_{1_j,\mathrm{max}}^G$, respectively, and following the same procedure used to derive $\lambda_{1-4}^{j}$ for the asymptotic SKR, the symplectic eigenvalues $\hat{\lambda}_{1-4}^{j}$ in { composable} finite-size SKR of GMCS can be obtained.

%
%


\end{document}